\documentclass{aa}  
\usepackage{bm}
\usepackage{natbib}
\usepackage{subcaption}
\usepackage{multicol}

\newcommand{\vect}[1]{\mathbf{#1}}
\renewcommand{\vec}{\vect}
\newcommand{\td}[2]{\frac{d #1}{d #2}}

\newcommand{\pd}[2]{\frac{\partial#1}{\partial#2}}

\newcommand{\pdd}[2]{\frac{\partial^2#1}{\partial#2^2}}

\usepackage{graphicx}
\usepackage{txfonts}
\begin{document}



\title{The effect of drifts on the decay phase of SEP events}

\titlerunning{Drifts during SEP events}        
   \author{N. Wijsen
          \inst{1,2}
          \and
          A. Aran\inst{2}
          \and
          B. Sanahuja\inst{2}
          \and
          J. Pomoell\inst{3}
          \and
          S. Poedts\inst{1,4}
          }

   \institute{Department of Mathematics/Centre for mathematical Plasma Astrophysics, KU Leuven, Belgium\\
              \email{nicolas.wijsen@kuleuven.be}
          \and
                 Departament F\'{i}sica Qu\`antica i Astrof\'{i}sica, Institut de Ci\`encies del Cosmos (ICCUB), Universitat de Barcelona (IEEC-UB), Spain
         \and
                 Department of Physics, University of Helsinki,  Finland
         \and
                Institute of Physics, University of Maria Curie-Sk{\l}odowska, Poland 
             }

   \date{Received:  30 October 2019; Accepted: 27 December 2019}

 
  \abstract
   {}
   {
   We study the effect of the magnetic gradient and curvature drifts on the pitch-angle dependent transport of solar energetic particles (SEPs) in the heliosphere, focussing on ${\sim}$3 -- 36 MeV protons. 
   By considering observers located at different positions in the heliosphere, we investigate how drifts may alter the measured intensity-time profiles and energy spectra.  
   We focus on the decay phase of solar energetic proton events in which a temporal invariant spectrum and disappearing spatial intensity gradients are often observed; a phenomenon known as the `reservoir effect' or the `SEP flood'.
   We study the effects of drifts by propagating particles both in nominal and non-nominal solar wind conditions.
   }
   {
   We used a three-dimensional (3D) particle transport model, solving the focused transport equation extended with the effect of particle drifts in the spatial term. 
   Nominal Parker solar wind configurations of different speeds and a magnetohydrodynamic (MHD) generated solar wind containing a corotating interaction region (CIR) were considered.
   The latter configuration gives rise to a magnetic bottle structure, with one bottleneck at the Sun and the other at the CIR. 
   We inject protons from a fixed source at 0.1 AU, the inner boundary of the MHD model.  
   }
   {
   When the drift induced  particle net-flux is zero,  the  modelled  intensity-time  profiles obtained at different radial distances along an IMF line show  the  same  intensity  fall-off after  the  prompt  phase  of  the particle event, which is in accordance with the SEP flood phenomenon. 
   However, observers magnetically connected close to the  edges  of  the particle injection site can experience, as a result of drifts, a sudden drop in the intensities occurring at different times for different energies such that no SEP flood phenomenon is established. 
  In the magnetic bottle structure, this effect is enhanced due to the presence of  magnetic field gradients strengthening the nominal particle drifts. 
   Moreover, anisotropies can be large for observers that only receive particles through drifts, illustrating the importance of pitch-angle dependent 3D particle modelling. 
   We observe that interplanetary cross-field diffusion can mitigate the effects of particle drifts.}
   {Particle drifts can substantially modify the decay phase of SEP events,  especially if the solar wind contains compression regions or shock waves where the drifts are enhanced. 
   This is, for example, the case for our CIR solar wind configuration generated with a 3D MHD model, where the effect of drifts is strong.
   A similar decay rate in different energy channels and for different observers requires the mitigation of the effect of drifts. 
   One way to accomplish this is through interplanetary cross-field diffusion, suggesting thus a way to determine a minimum value for the cross-field diffusion strength. }


   \maketitle
%

\section{Introduction}

When solar energetic particles (SEPs) are released from their acceleration site, they commence a journey through the inner heliosphere and beyond. 
Being charged particles, SEPs are subject to the Lorentz-force, gyrating along the interplanetary magnetic field (IMF) lines embedded in the solar wind. 
The IMF contains turbulent fluctuations that scatter energetic particles and possibly transport them perpendicular to the magnetic field. 
In addition, gradients in the IMF and the curvature of the IMF lines induce particle drifts, moving particles across IMF lines. 

Recently, \cite{marsh13} and \cite{dalla15,dalla17} used full-orbit test particle simulations to model particle propagation in interplanetary space.  
These authors illustrate that drifts can have a major effect on the distribution of SEPs in the heliosphere, especially when considering high-energy particles (> 100 MeV protons) or ions with a high mass-to-charge ratio.
The importance of drifts also increases towards higher heliocentric latitudes as discussed analytically, for example, by \cite{dalla13}. 
In addition, \cite{dalla15} show how drifts in the opposite direction of the convective electric field in the solar wind result in the deceleration of energetic particles. 
This process has also been discussed previously in \cite{leRoux07} and \cite{leRoux09} where the authors instead used a kinetic transport equation approach. 

In the past, the effect of particle drifts has mainly been studied in the context of cosmic rays (CRs) as drifts provide a natural explanation for the observed heliospheric CR modulation over the solar cycle \citep{jokipii77} and for different features in the energy spectrum of cosmic-rays, such as the first and second knees \citep{ptuskin93}.
Modelling CR transport in space is typically  done by solving the Parker equation \citep{parker65}, which describes the evolution of an isotropic distribution function. 
The spatial evolution is described through a diffusion process, in which the effects of particle drifts  enter via the off-diagonal elements of the corresponding diffusion tensor. 
However, in order to explain the observations of CR modulation using the Parker equation approach, the effects of particle drifts are often too strong and hence have to be reduced \citep[e.g.][]{potgieter89}.
Full-orbit test particle simulations in a prescribed turbulent magnetic field configuration have indeed illustrated that the drift coefficients in the Parker equation can be suppressed if the level of turbulence is high enough \citep[e.g.][]{giacalone99,candia04,minnie07,tautz12,engelbrecht19}. 
However, it is important to note that the Parker equation is only valid for isotropic particle distributions, which is not necessarily valid for SEPs as they are often characterised by strong anisotropies \citep[e.g.][]{heras94}. 
For such events, the focused transport equation (FTE) \citep[e.g.][]{roelof69} is more appropriate to describe the particle distributions. 
It is not clear how the results on drift reduction translate to the situation of a pitch-angle dependent distribution function, especially because the particle drifts are in that case also pitch-angle dependent.
Assuming a large-scale Parker spiral magnetic field configuration \citep{parker58}, \cite{marsh13} use their  full-orbit test particle model to show that the scattering conditions have only a weak effect on the particle drifts. 
However, it is not clear if this is due to the high energies they considered, the relatively large parallel mean free paths, or the way their model treats the effects of turbulence through random scattering
events that are Poisson-distributed in time.

Aside from the work mentioned above, particle drifts have been largely  ignored when modelling SEP transport in the inner heliosphere.
Instead, much more attention has been devoted to cross-field diffusion,  as it is perceived as an efficient manner to spread particles in the heliosphere. 
In several cross-field diffusion models, the perpendicular transport of energetic particles is a result of particles diffusing along magnetic field lines that undergo a random walk \citep[see e.g.][and references therein]{shalchi09b}. 
Cross-field diffusion is most effective where there are large particle intensity gradients perpendicular to the magnetic field \citep[see e.g.\ Eq.~(6) in][]{wijsen19c}, hence, reducing these intensity gradients. 
Therefore, cross-field diffusion provides a tentative explanation for multi-spacecraft observations that have shown that particle intensities measured in the decay phase of large SEP events by widely separated spacecraft often evolve similarly in time \citep{mcKibben72}. In addition, the decay phases of these SEP events often have energy spectra that are invariant both in space and time \citep[e.g.][]{reames97a,reames97b}.  
 \cite{roelof92} suggested that diffusion barriers produced by coronal mass ejections (CMEs) and interplanetary shocks may be the responsible mechanism and, hence, they coined the term  'reservoir phenomenon'.
 \cite{reames96} proposed that the formation of the reservoir is based on the trapping of particles in a slowly expanding bottle behind a  CME. However, it has been noted by authors such as \cite{wang15} 
that such an expanding magnetic bottle model would still need
cross-field diffusion to  reduce the spatial gradients of SEP fluxes. 
In addition, for some SEP events, 
CMEs are not directly observed by the spacecraft, yet the
reservoir phenomenon is still observed \citep{dalla03} and in some cases, the reservoir phenomenon commences before the arrival of the CME \citep{reames99,wang15}. It is for this reason, among others, that \cite{he17a} renamed the reservoir phenomenon instead as the 'SEP flood'. In these studies on the reservoir or flood phenomenon, the effects of drifts appear to have been ignored. 
This is in spite of the observation that during the decay phase, particles will have had enough time to drift a substantial distance. In addition, large scale structures in the IMF, such as magnetic bottles, may produce enhanced particle drifts due to, for example, the presence of strong magnetic field gradients.  

In this work, we study the effect of particle drifts, which are due to the gradient and curvature of the magnetic field, on the energy spectra and decay rates of SEP time-intensity profiles as measured by observers located at different positions in the inner heliosphere. We assume different solar wind configurations, including nominal Parker solar winds of 300~km~s$^{-1}$ and 700~km~s$^{-1}$ and a solar wind containing a corotating interaction region (CIR), modelled with the three-dimensional (3D) MHD model EUHFORIA \citep{pomoell18}. Particles are propagated under different scattering conditions and assuming different injection profiles. 
We find that the energy dependence of drifts might have a substantial effect in some cases, especially for observers magnetically connected to the edges of the SEP injection sites. Moreover, the  magnetic bottle that is naturally formed in a solar wind containing a CIR, produces strong drifts making particle intensities measured in different energy channels to decay at different rates. 

The paper is structured as follows. In Section~\ref{sec:modelling}, we briefly describe our particle transport model. In Sections~\ref{sec:Imf_obs} and~\ref{sec:Parker_lat}, we study the effect of drifts on the intensity-time profiles measured by observers placed along different IMF lines in a Parker configuration with solar wind speeds of 300~km~s$^{-1}$ and 700~km~s$^{-1}$. In Section~\ref{sec:decay}, we compare the intensity decay rate of different energy channels. In Sections~\ref{sec:erg_peak} and~\ref{sec:erg_fluence}, we show how drifts affect the measured particle energy spectra, obtained from the peak intensities and the fluence, respectively. The effect of cross-field diffusion on our results is discussed in Section~\ref{sec:cfd}. In Section~\ref{sec:cir}, we investigate the particle drifts in a solar wind containing a CIR. We conclude with a summary and discussion in Section~\ref{sec:summary}.

\section{Modelling SEP transport}\label{sec:modelling}
To study SEPs in the inner heliosphere, we model the evolution of the directional particle intensity
$j(\vec{x},p,\mu,t)$
using the focused transport equation (FTE), extending it with particle drifts in the spatial term \citep[see e.g.][]{leRoux09}:
\begin{equation}\label{eq:fte}
\begin{aligned}
\pd{j}{t} &+\pd{}{\vec{x}}\cdot\left[\left(\td{\vec{x}}{t}+\pd{}{\vec{x}}\cdot \bm{\kappa}_\perp\right)j\right]+\pd{}{\mu}\left[\left(\td{\mu}{t}+ \pd{D_{\mu\mu}}{\mu}\right) j\right] \\ &+ \pd{}{p}\left(\td{p}{t}j\right) \\ =& \pdd{}{\mu}\left[D_{\mu\mu}j\right] + \pd{}{\vec{x}}\cdot\left[\pd{}{\vec{x}}\cdot\left(\bm{\kappa}_\perp j\right)\right],
\end{aligned}
\end{equation}
with
\begin{eqnarray}
\td{\vec{x}}{t} &= &
\vec{V}_{\rm sw}+ \vec{V}_d+\mu \varv\vec{b} \label{eq:fte_x}\\
\td{\mu}{t}&=&
\frac{1-\mu^2}{2}\Bigg(\varv \nabla\cdot\vec{b} + \mu \nabla\cdot\vec{\vec{V}_{\rm sw}} - 3 \mu \vec{b}\vec{b}:\nabla\vec{\vec{V}_{\rm sw}}\label{eq:fte_mu}\\
\notag 
&&- \frac{2}{\varv}\vec{b}\cdot\td{\vec{\vec{V}_{\rm sw}}}{t} \Bigg) \\
\td{p}{t} &=&
 \Bigg( \frac{1-3\mu^2}{2}(\vec{b}\vec{b}:\nabla\vec{\vec{V}_{\rm sw}}) - \frac{1-\mu^2}{2}\nabla\cdot\vec{\vec{V}_{\rm sw}}\label{eq:fte_p} \\
\notag 
 && -\frac{\mu }{\varv}\vec{b}\cdot\td{\vec{\vec{V}_{\rm sw}}}{t}\Bigg) p.
\end{eqnarray}
In these equations, 
$\vec{x}$ denotes the phase-space spatial coordinate and $t$ the time, both measured in an inertial frame, whereas the cosine of the pitch angle $\mu$ and the momentum magnitude $p$ or speed $\varv$ are expressed in a frame that is comoving with the solar wind. 
Furthermore, $\vec{V}_{\rm sw}$ is the solar wind velocity, and
$\vec{b}$ the unit vector in the direction of the mean magnetic field,
$D_{\mu\mu}$ is the pitch-angle diffusion coefficient, 
$\bm{\kappa}_\perp$ the spatial cross-field diffusion tensor, and
$\vec{V}_d$ the pitch-angle dependent drift velocity due to the gradient and curvature of the mean magnetic field, that is,
\begin{equation}\label{eq:drift}
\vec{V}_d = \frac{\varv p}{QB}\left[\frac{1-\mu^2}{2} \left( (\nabla\times \vec{b})_\parallel + \frac{\vec{b}\times\nabla B}{B}\right) + \mu^2 (\nabla\times \vec{b})_\perp \right],
\end{equation}
where $Q$ is the particle charge, and the subscripts $\parallel$ and $\perp$ denote, respectively, the parallel and perpendicular components with respect to the magnetic field $\vec{B}$. 
As illustrated in the appendix of \cite{leRoux09}, the FTE includes the effects of the gradient and curvature drift on the  particle energy changes by the convective electric field. 
In addition, \cite{leRoux09} show that by including the  gradient and curvature drifts in the spatial term of the FTE, the equation becomes equivalent to the standard drift guiding kinetic equation \citep{littlejohn83}. 

\begin{table}
\caption{Energy channels and their geometric mean $\langle E \rangle$ in MeV.}            
\label{table:energy}      
\centering                            
\begin{tabular}{c c }          
\hline\hline    
$E_{\rm min} - E_{\rm max}$ & $\langle E \rangle$   \\
\hline
$\phantom{0}2.39 - \phantom{0}3.06 $& 2.70 \\
$\phantom{0}3.06 - \phantom{0}3.91 $& 3.46 \\
$\phantom{0}3.91 - \phantom{0}5.00 $& 4.42 \\
$\phantom{0}5.00 - \phantom{0}6.39 $& 5.65 \\
$\phantom{0}6.39 - \phantom{0}8.18 $& 7.23 \\
$\phantom{0}8.18 - 10.46 $& 9.25 \\
$10.46 - 13.37 $& 11.82 \\
$13.37 - 17.10$& 15.12 \\
$17.10 - 21.87 $& 19.34\\
$21.87 - 27.96 $& 24.73\\
$27.96 - 35.76 $& 31.62\\
\hline 
\end{tabular}
\end{table}

Our particle transport model solves Eq.~\eqref{eq:fte} by integrating the equivalent set of It\^o stochastic differential equations forward in time \citep{gardiner04}.  
For details regarding the numerical procedures used by our model, we refer to \cite{wijsen19a}. 
In our simulations, the particles are sampled in the phase-space volume $2\pi r^2 \sin(\theta)\Delta \vartheta \Delta \varphi  \Delta r \Delta E \Delta \mu$, where $E$ is the particle energy and $(r,\vartheta,\varphi)$ are spherical coordinates associated to the inertial frame. 
We choose $\Delta r = 0.01$ AU, $\Delta \vartheta = \Delta \varphi = 0.5^\circ$, $\Delta \mu = 0.1$, $\log_{10}(\Delta E) = 0.1068215$. 
The energy channels that we consider in this work are listed in Table~\ref{table:energy} and in the following; when discussing a specific energy channel, we simply refer to the geometric mean $\langle E \rangle$ of the channel.

All simulations discussed in this work are performed using $1.5\times 10^8$ particles. To increase the statistics further, we  average over a time period of 5 minutes.

To mimic a prompt injection of particles, the protons are injected according to a Reid-Axford time profile \citep{reid64} and a power law in energy $E$:
\begin{equation}\label{eq:RA}
 j(t) \propto \frac{E^{-\gamma}}{t}\exp{\left(-\frac{\beta}{t} - \frac{t}{\tau}\right)},   
\end{equation}
where we choose the parameters of $\beta = 0.2 $~hours, $\tau=1$~hours and $\gamma = 3$, unless stated otherwise. We inject particles with an energy range between 2 and 60 MeV uniformly from a region that  spans $5^\circ$ in both longitude and latitude, located at 0.1 AU and centred on the solar equatorial plane. The upper energy of 60 MeV is 24.24 MeV above the highest energy channel considered in our simulations (see  Table~\ref{table:energy}). This is done to prevent adiabatic deceleration from depleting the highest energy channel, hence allowing us to focus on the effects of drifts more easily. To obtain comparable  statistics in all energy channels, our model injects particles uniformly between 2 and 60 MeV. However, upon injection all particles receive a statistical weight derived from the power law distribution, which is taken into account when sampling the particles.

The particle scattering conditions are determined by the pitch-angle diffusion coefficient $D_{\mu\mu}$ for which we use quasi-linear theory to prescribe its functional form \citep[see][for details]{wijsen19a}. 
The level of particle diffusion is quantified by assuming a constant radial mean free path $\lambda_\parallel^r$ \citep{bieber94}. 
In this work, we choose either $\lambda^r_\parallel = 0.3$~AU or  $\lambda_\parallel^r=0.08$~AU. 
We do not consider cross-field diffusion, that is, $\kappa_\perp = 0$, except in Section~\ref{sec:cfd}, where we describe the cross-field diffusion model. 
We inject particles in a Parker IMF of positive polarity for slow 300~km~s$^{-1}$ and fast 700~km~s$^{-1}$ solar wind configurations. 
Both solar winds are computed assuming a constant sidereal solar rotation period of 25.4 days and a radial magnetic field strength of $1.85$~nT at 1.0 AU. 
In addition, we inject particles in a solar wind configuration containing a CIR, which is generated by a 3D ideal MHD model. 
The set-up of these latter simulations is discussed in more detail in Section~\ref{sec:cir}.

Furthermore, in the results presented below, we focus on observers corotating with the IMF, since the effects of corotation have already been discussed thoroughly in previous studies \citep[see e.g.][]{droge10,giacalone12,wijsen19a}.
Finally, in this work we focus on the effect of drifts in the latitudinal direction only since, close to the equatorial plane,  the latitudinal component of the particle drift velocity is dominant with respect to the azimuthal component  \citep[e.g.][]{dalla13}. Therefore, in the following we consider observers at different latitudes but they are always  magnetically connected to the longitudinal centre of the injection region. 

\section{Results}
\subsection{Observers along IMF lines}\label{sec:Imf_obs}
\begin{figure*}
        \centering
        \begin{tabular}{cc}
        \includegraphics[width=0.4\textwidth]{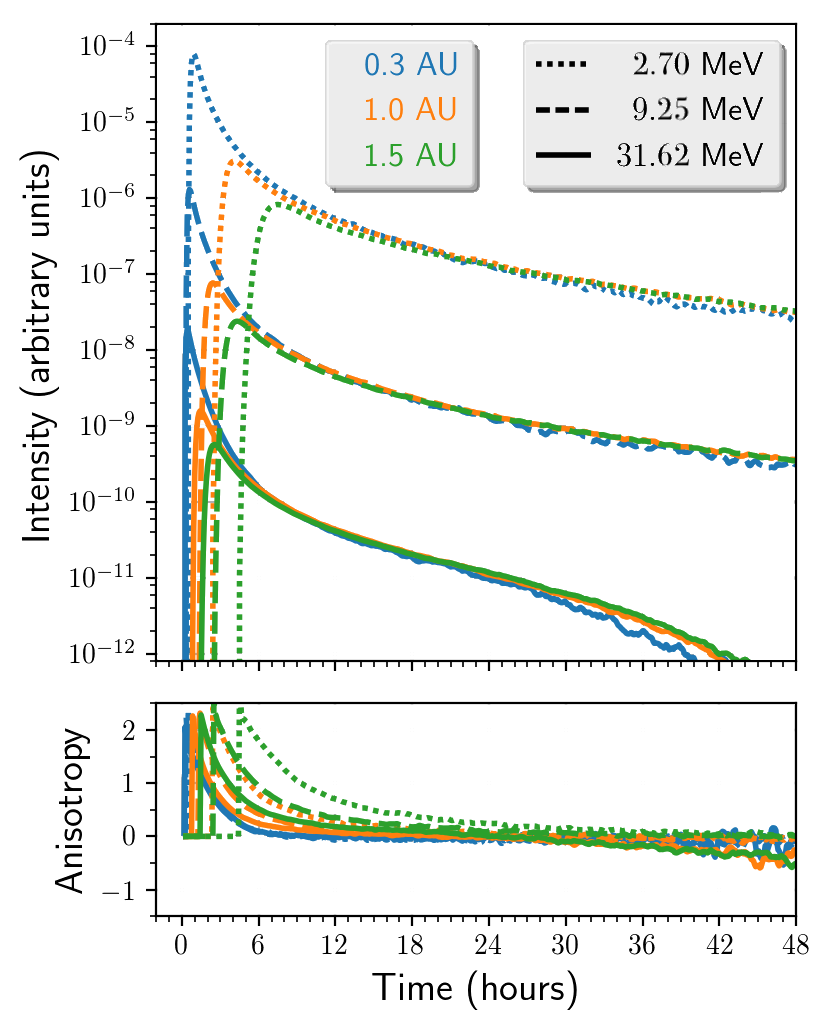} & 
        \includegraphics[width=0.4\textwidth]{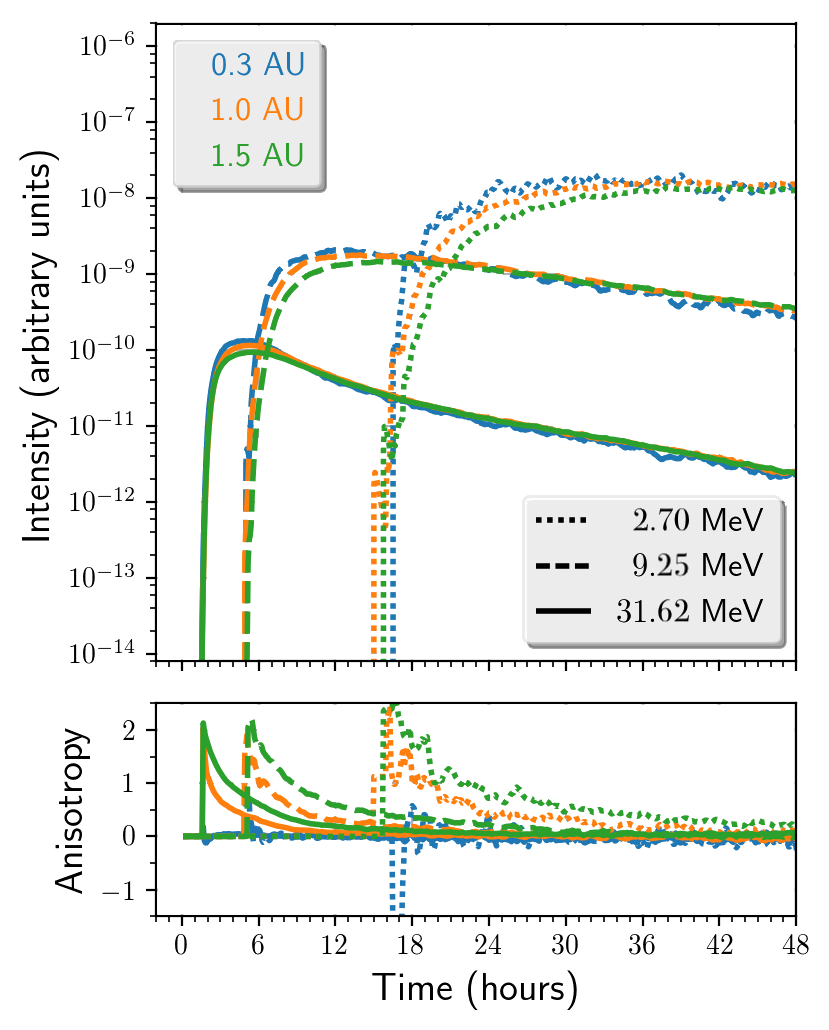} \\ 
        \includegraphics[width=0.45\textwidth]{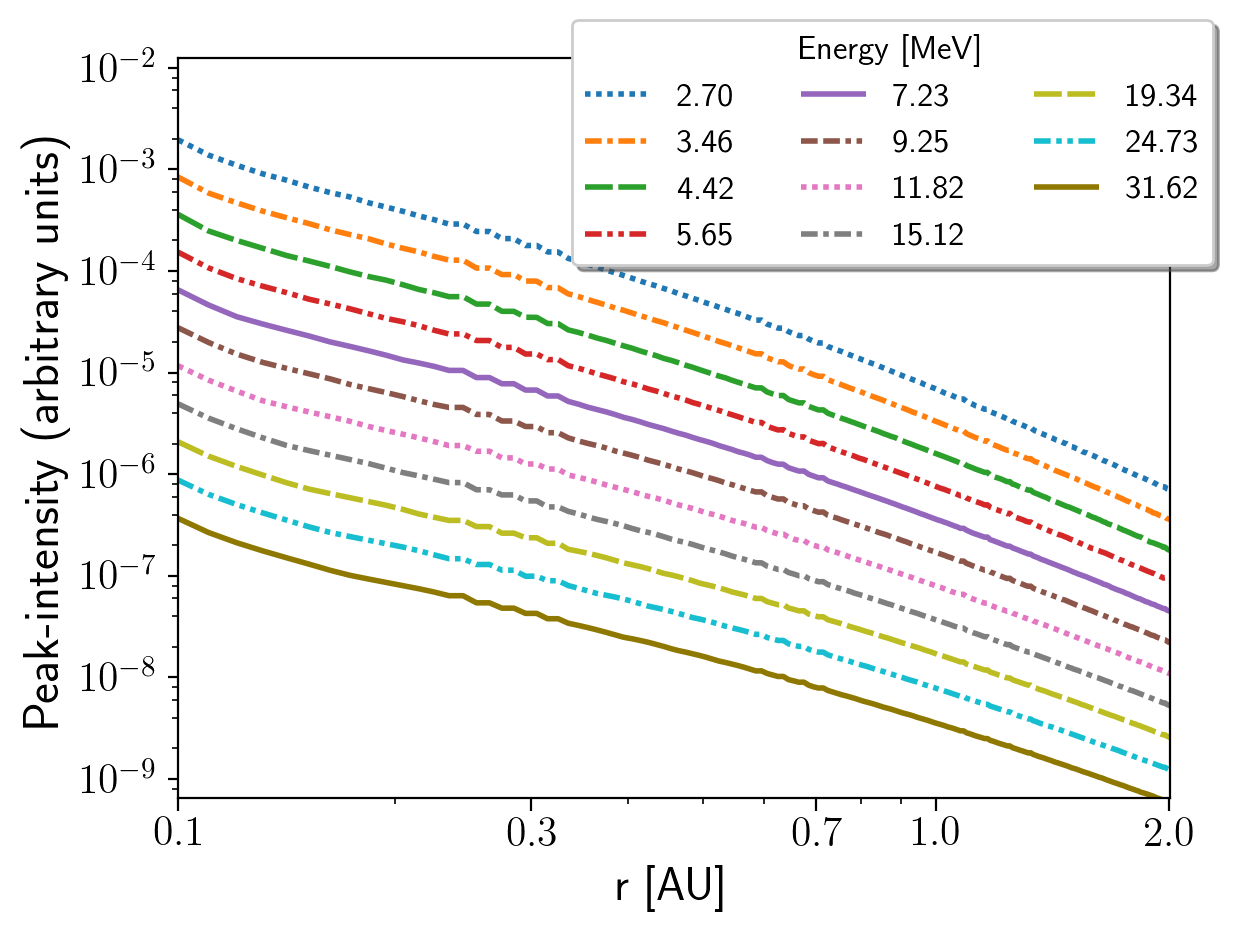} & 
        \includegraphics[width=0.45\textwidth]{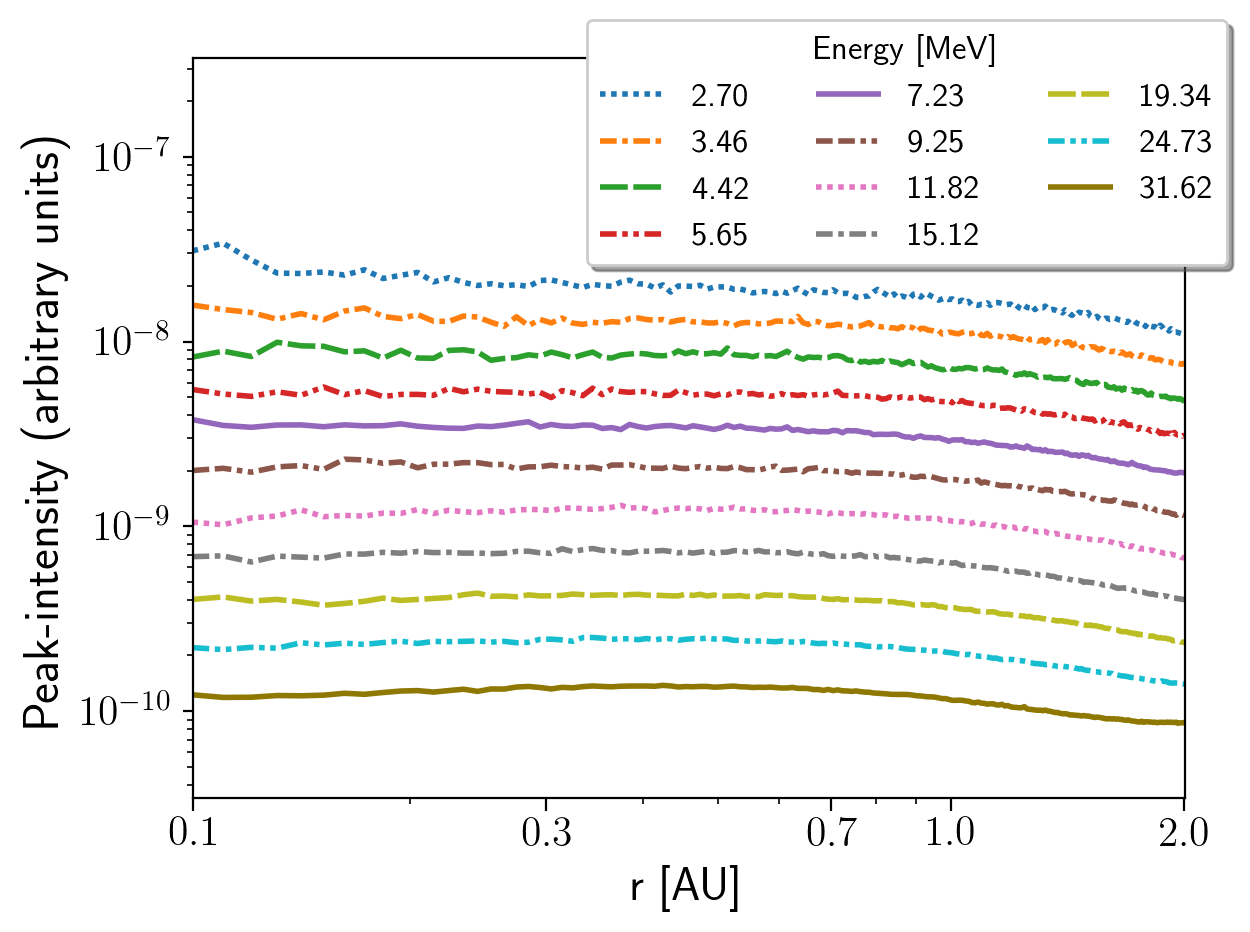}
    \end{tabular}
    \caption{\textit{Upper row:}   intensity (top) and anisotropy (bottom)-time profiles for three observers located along an IMF line in a 300~km~s$^{-1}$ Parker solar wind rooted at 0.1 AU at $0^\circ$ latitude (left panel) and at $-3^\circ$ latitude (right panel). Both panels show the intensity-time profiles for observers at a heliocentric radial distance of 0.3 AU (blue), 1 AU (orange), and 1.5 AU (green), and for protons of 2.70~MeV (dotted curves), 9.25~MeV (dashed) and 31.62~MeV (solid). \textit{Lower row:} peak-intensities of the simulated proton channels (colour coded) as a function of radial distance for an IMF line at $0^\circ$ latitude (left panel) and at $-3^\circ$ latitude (right panel). }
     \label{fig:3obsImf300}
\end{figure*}
In this section, we study the intensity-time profiles measured by observers placed at heliocentric radial distances of 0.3 AU, 1.0~AU, and 1.5 AU for two different IMF lines. 
One IMF line has its foot-point located in the centre of the particle injection region and is hence referred to as the 'central IMF line'. The other has its foot-point $3^\circ$ more towards the south, that is, $0.5^\circ$ south of the southernmost edge of the injection region and is therefore referred to as the 'southern IMF line'. 
The latter magnetic field line is  not connected, thus, to the injection region at 0.1 AU and so, any particle intensities measured along this IMF line can solely be attributed to drifts since the cross-field diffusion is set to zero. 

The results shown in Fig.~\ref{fig:3obsImf300} refer to particles experiencing a constant radial mean free path of 0.3 AU in a 300~km~s$^{-1}$ Parker solar wind.  The upper left panel of Fig.~\ref{fig:3obsImf300} shows the intensity-time profiles in three different energy channels for the observers located along the central IMF line.
After the prompt onset, particle intensities evolve for the three observers identically, that is, the SEP flood phenomenon is reproduced. As can be seen, the corresponding anisotropies indicate at the onset of the particle event, a clear anti-sunward net-flux of particle, whereas during the decay phase, the particle distribution isotropizes.  

The upper right panel of Fig.~\ref{fig:3obsImf300} shows the intensity-time profiles for observers along the IMF line not connected to the injection region. 
Since magnetic gradient and curvature drifts are both directed southwards in a Parker spiral of positive polarity \citep[see e.g.][]{dalla13}, particles eventually reach this IMF line.
However, this process requires time, explaining why the prompt phase of the particle event is not apparent in the intensity-time profiles.
As can be seen, the onset time of the particle event and the attained intensities is approximately identical for the three observers. 
For the $2.70$ MeV energy channel, the onset time is delayed up to ${\sim}16$ hours. 
Meanwhile, at this time, the SEP flood phenomenon has already been established for a well-connected observer and, hence, it is also observed for the southwards drifting particles. 
This is the reason why  the onset time and the attained intensities are approximately identical along the southern IMF line. 
The significant difference between the arrival time of particles residing in the different  energy channels is due to the  proportionality of the drifts to the particle energy.

Finally, in looking at the anisotropies \citep[see e.g.][for the definition]{wijsen19a}, we see that at the onset time, the observers at 1.0~and 1.5 AU see strong positive anisotropies corresponding to anti-sunward streaming protons. 
This is the case even for the $2.70$~MeV channel, which sees the particle onset only after 16~hours.
At this time, the anisotropies in this channel are much smaller for the same observers along the central IMF line. 
It is only for the observer at 0.3 AU that the particle distribution already become isotropic at the time of the onset. 

The second row of Fig.~\ref{fig:3obsImf300} shows the peak intensity along both IMF lines as a function of the heliocentric radial distance for various energy channels. 
For the well-connected IMF line, we see a decreasing trend resulting from the expanding magnetic flux tubes and  particle scattering, as studied previously in, for example,\ \cite{lario07} and \cite{he17b}. 
The peak intensities along this IMF line track the prompt phase of the particle event. 
For the southern IMF line, we see that the peak intensity is approximately constant with radial distance. 
This is because  the peak intensities along the IMF line are not reached during the prompt phase of the particle event but during the decay phase when the SEP flood phenomenon is established.  

\begin{figure*}
        \centering
        \begin{tabular}{ccc}
        \includegraphics[width=0.3\textwidth]{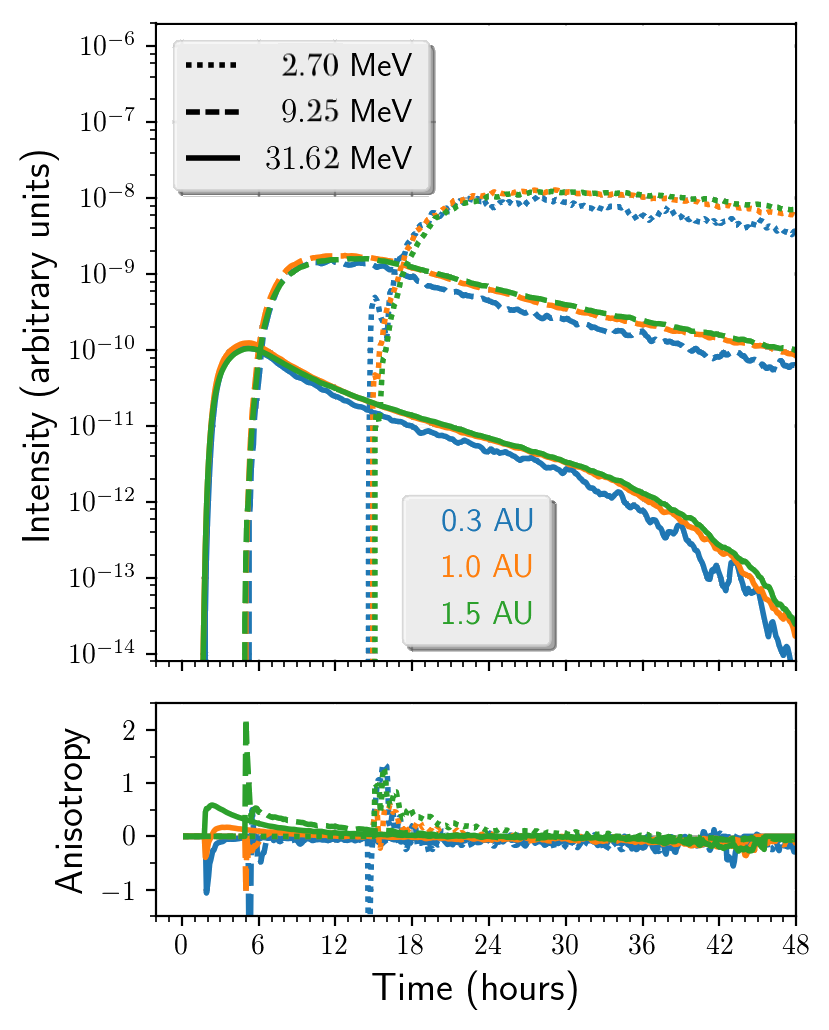} & 
        \includegraphics[width=0.3\textwidth]{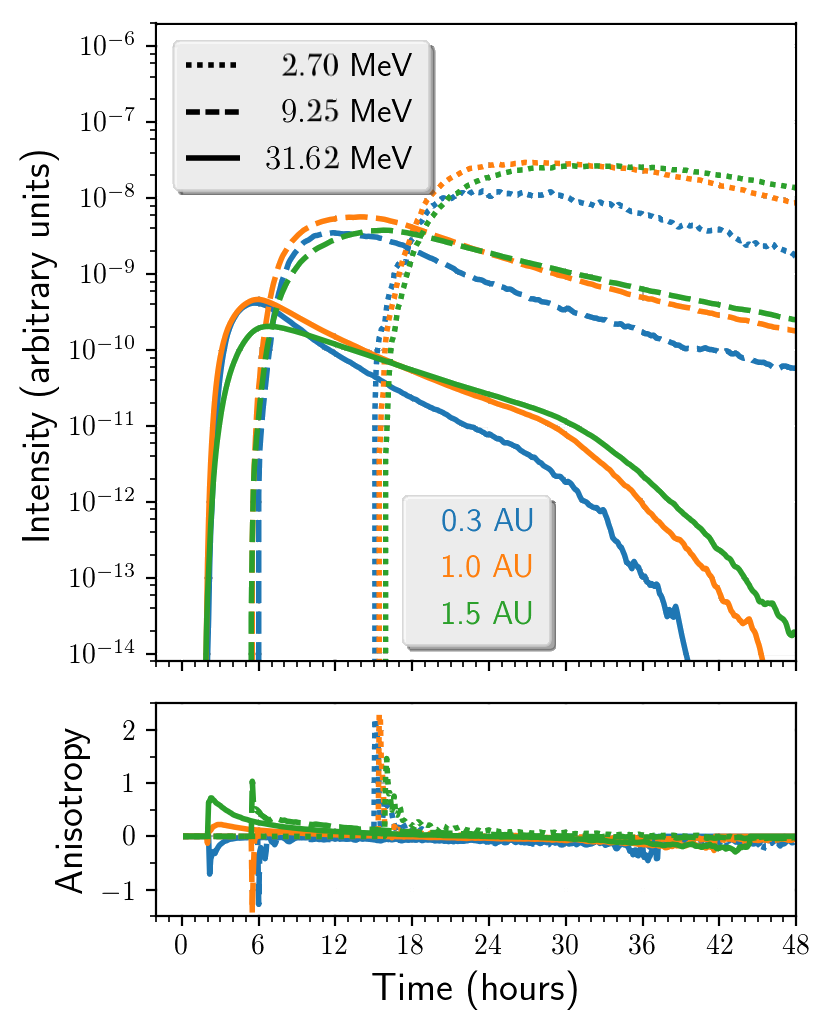} &
        \includegraphics[width=0.3\textwidth]{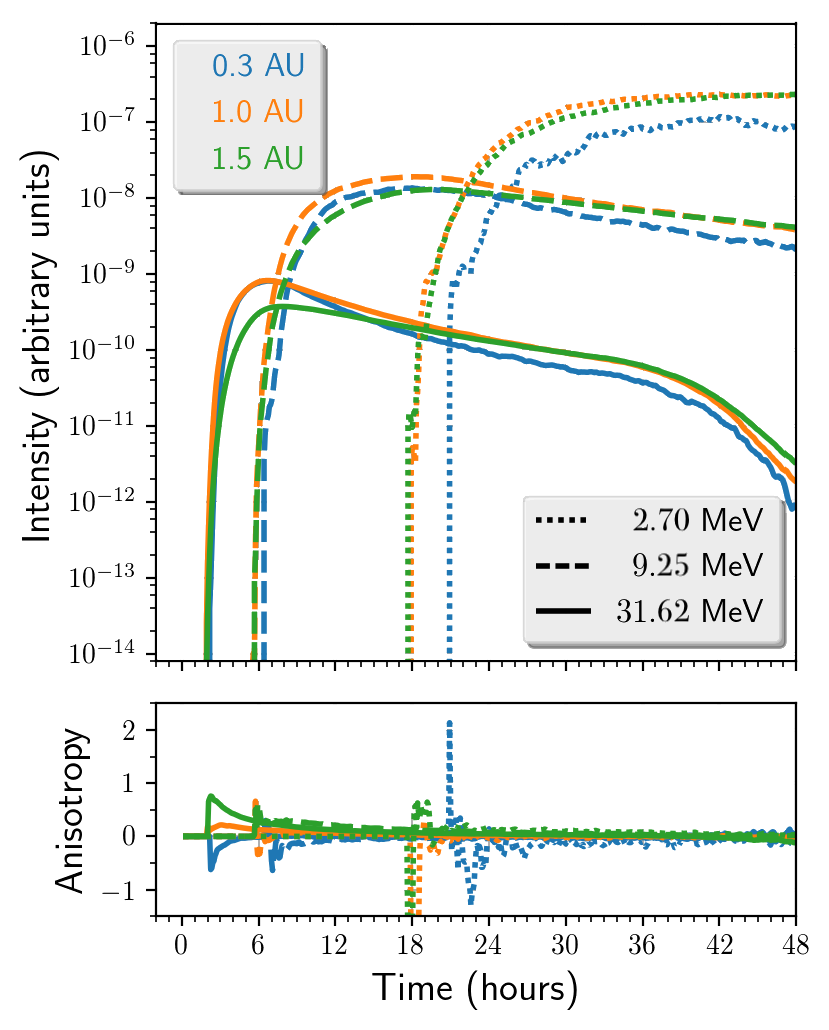} 
    \end{tabular}
    \caption{ Intensity-time profiles for an  observer at  $-3^\circ$ latitude, in a 700~km~s$^{-1}$ Paker solar wind. Left panel: $\lambda_\parallel^r = 0.3$~AU. Central panel:  $\lambda_\parallel^r = 0.08$~AU. Right panel:  $\lambda_\parallel^r = 0.08$~AU and with particle deceleration switched off.   }
     \label{fig:3obsImf700}
\end{figure*}

To give a better idea of the effect of the solar wind speed on the particle drifts, Fig.~\ref{fig:3obsImf700} shows the intensity-time profiles for the case of a solar wind of 700~km~s$^{-1}$ instead of 300~km~s$^{-1}$ as previously shown.  
The left panel of the figure shows the intensity-time profiles for the observers along the southern IMF line, where we assumed the same scattering conditions as before (i.e. $\lambda_\parallel^r= 0.3$ AU).
Thus, the only difference in the simulation parameters between this panel and the upper right panel of Fig.~\ref{fig:3obsImf300} is the solar wind speed.  
As can be seen, similar intensities are achieved in both cases, yet the decay rate of the $31.62$ MeV channel is considerably higher in the 700~km~s$^{-1}$ case, especially towards the end of the simulation.
Moreover, the flood phenomenon is not as well established in the 700~km~s$^{-1}$ solar wind since the observer at 0.3~AU measures slightly lower particle intensities than the observers at 1.0~and 1.5~AU. 
A remarkable difference between the  slow and fast solar wind cases is that the anisotropies in the former are significantly larger during the first hours following the particle onset. 
In addition, in contrast to the slow solar wind case, both at 0.3~AU and 1.0~AU in the fast solar wind case, the observers measure negative anisotropies in the $9.25$~MeV and  the $31.62$~MeV energy channels, indicating sun-ward streaming protons. 
From the expression of the drift velocities in a Parker spiral \citep[see Eqs.~(23) -- (29) in][]{dalla13}, it can be derived that for the assumed solar wind configurations,  the drifts in the latitudinal direction are larger in the fast than in the slow solar wind for heliocentric distances $r \gtrsim 0.8$~AU. 
In contrast, for $r \lesssim 0.8$~AU, the slow solar wind contains the larger particle latitudinal drift speeds. 
These radial dependencies of the drifts lead to the different anisotropies measured in both solar wind configurations.

Next, we examine the effect of the radial mean free path on the particle drifts. 
The central panel of Fig.~\ref{fig:3obsImf700} shows the intensity-time  profiles for $\lambda_\parallel^r = 0.08$~AU. 
As can be seen, decreasing the radial mean free path increases the obtained intensities but also the decay rate, especially for the observer at 0.3 AU.  
Despite the smaller mean free path and hence the stronger diffusion along the IMF line, the flood phenomenon is less established compared to the previous simulation. 
This result can be attributed to the effects of particle deceleration.  
The smaller mean free path traps particles a longer time at small radial distances, where adiabatic deceleration is strongest. 
To illustrate this, we show in the right panel of Fig.~\ref{fig:3obsImf700} the intensity-time  profiles for a simulation where the terms causing particle deceleration ( i.e. Eq.~\eqref{eq:fte_p}) were switched off. 
In this case, the different observers do obtain largely similar intensities in the corresponding energy channels. We also note that the onset of the particle intensities in the $2.70$ MeV energy channel is delayed by about 2 hours.
This is because in the simulation including particle deceleration, particles with higher injection energy and hence larger initial drift velocity  end up populating the $2.70$ MeV energy channel, therefore, arriving earlier. This  illustrates the important combined effect of particle deceleration and particle drifts. 

Finally, it is worth noting that the peak-intensities of the drifted particles are larger for the smaller particle mean free path case. This in contrast to what is obtained for well-connected observers, where  peak-intensities increase for larger mean free paths \citep{lario07, agueda12}. Transport under small mean free paths favours the spreading of particles via drifts, translating into enhanced particle intensity levels observed all along the southern IMF line, whereas in the case of the central IMF line, the less anisotropic streaming conditions of the particles translate into larger peak intensities for the observers closer to the Sun with respect to those attained at further distances \citep{lario07}. 

\subsection{Observers at different latitudes}\label{sec:Parker_lat}
\begin{figure*}
        \centering
        \begin{tabular}{cc}
        \includegraphics[height=0.59\textwidth]{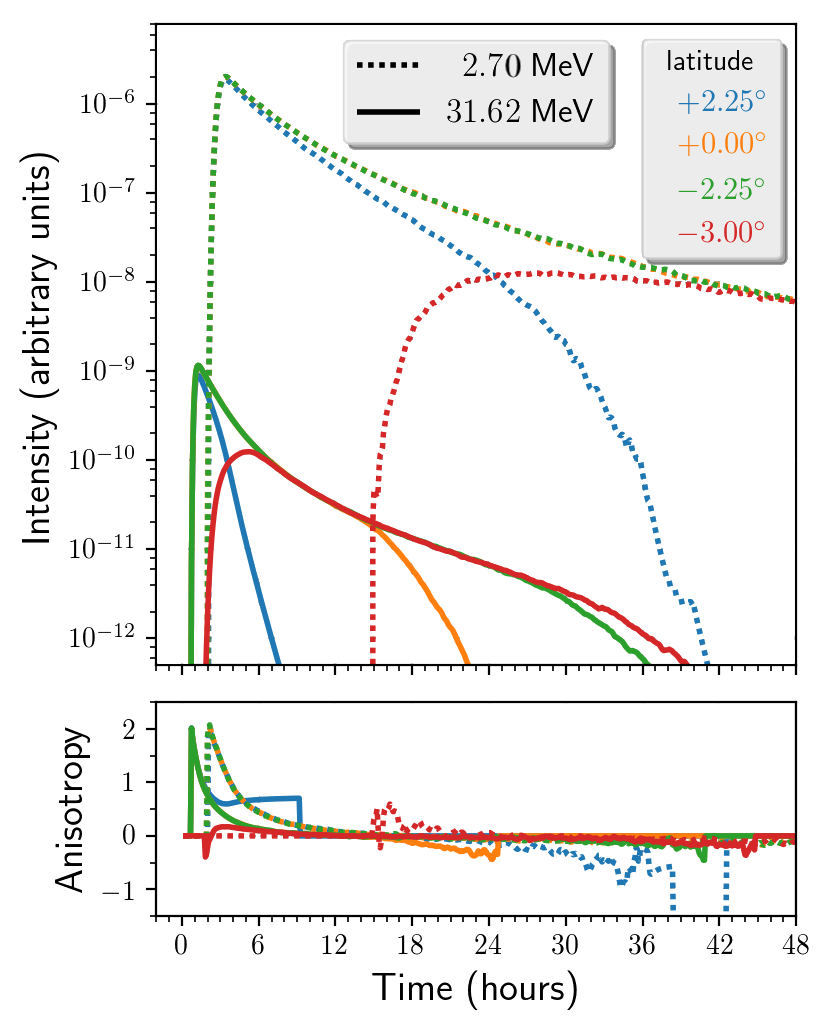} & 
        \includegraphics[height=0.59\textwidth]{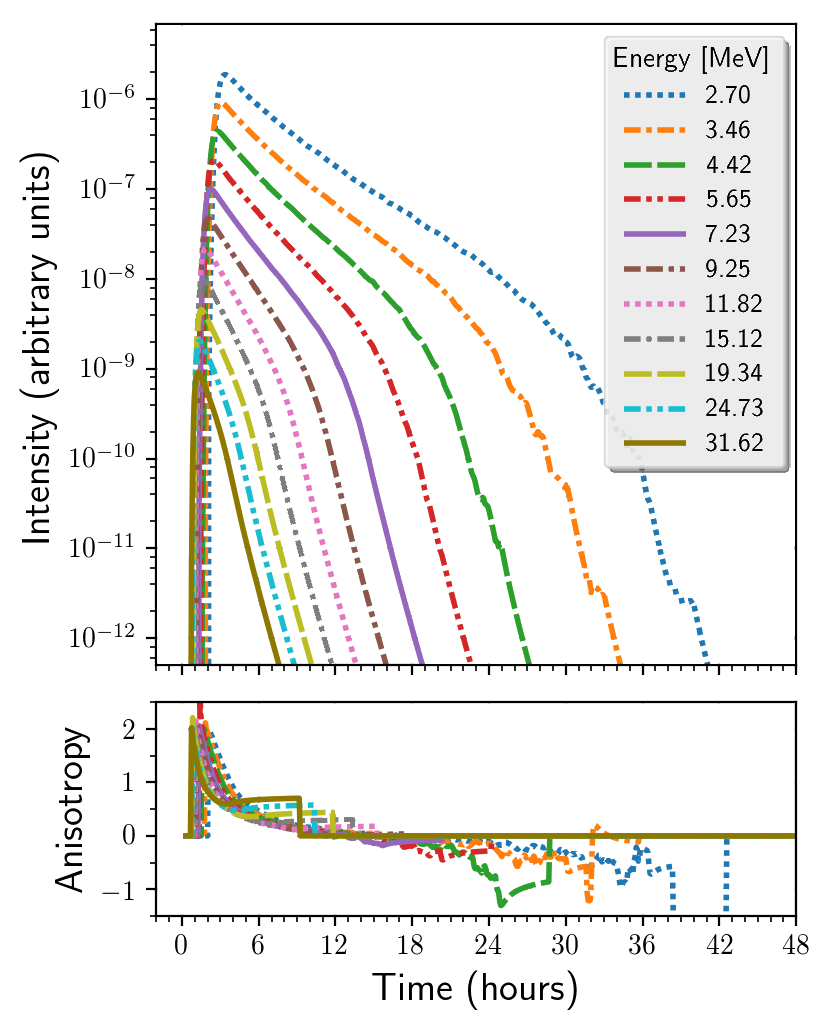}
    \end{tabular}
    \caption{\textit{Left panel:} 2.70~MeV and 31.62~MeV proton intensity-time profiles for observers located at 1.0~AU and at different latitudes, as indicated by the insets. \textit{Right panel:} intensity-time profiles with a complete energy coverage for an observer located a 1.0~AU and $2.25^\circ$ latitude (depicted by blue curves in the left panel).  }
     \label{fig:4obsClt700}
\end{figure*}

In this section, we study in detail the intensity-time profiles of corotating observers located at 1.0~AU, but at different latitudes. Specifically, we consider observers at latitudes $2.25^\circ$, $0^\circ$, $-2.25^\circ$, and $-3^\circ$. 
These latitudes denote the centre of the corresponding sampling box, which has a latitudinal width $\Delta\vartheta = 0.5^\circ$ (see Section~\ref{sec:modelling}).
Only the first three observers are thus magnetically connected to the particle injection region.
A 700~km~s$^{-1}$ solar wind and a $\lambda_\parallel^r = 0.3$~AU are assumed. 
The left panel of Fig.~\ref{fig:4obsClt700} shows the intensities and anisotropies for a high-energy (31.62 MeV) and a low-energy (2.70~MeV) channel. In these figures, the anisotropy is set to zero when the corresponding intensity drops below a threshold value of $10^{-13}$ to avoid spurious, non-significant fluctuations of the anisotropies. 
The intensity at high energy is observed to drop quickly for the observer located at $2.25^\circ$ latitude (blue curves). 
This observer is connected to a point close to the northern edge of the injection region and since the particles experience   drifts that are solely southward in a Parker IMF of positive polarity, particles are rapidly drained from this field line due to the drifts. The right panel of Fig.~\ref{fig:4obsClt700} displays  the intensity-time profiles for all eleven energy channels for this same observer.
In this panel we see clearly see how the proportionality of the drifts with energy of the particles makes the higher-energy channels drop to zero first.

It is important to note that the drops in intensity in  Fig.~\ref{fig:4obsClt700} are influenced by the assumed spatial sampling resolution. 
If, for example,  the latitudinal extent of the sampling box would be reduced, the southern edge of the sampling box would be shifted northward, (i.e. connected even closer to the northern edge of the injection region) and, hence, the drop-offs in intensity would occur earlier. 

The solid orange curve in the left panel of Fig.~\ref{fig:4obsClt700} shows that for the observer located in the equatorial plane, the intensities in the highest energy channel eventually also drop to zero. 
However, this occurs later than for the $2.25^{\circ}$ observer because during the first hours of the particle event, the observer at $0^\circ$ latitude receives drifting particles that were injected north of the solar equatorial plane; the net southward particle flux due to drifts is thus initially zero. 
With time, all high-energy particles drift southwards of the observer and the in-flux of these particles drops to zero, resulting in the increased decay rate.
This does not happen for the $2.70$ MeV channel in the depicted time-range, since the particle drifts are much slower at the low energies. 
Similarly, for the observer located at $-2.25^\circ$ latitude (green curves in the left panel of Fig.~\ref{fig:4obsClt700}), the increase in the decay rate of the $31.62$ MeV channel is only seen after ${\sim} 28$ hours.

As discussed in the previous section, the observer located at $-3^\circ$ latitude (red curves in the left panel of Fig.~\ref{fig:4obsClt700}) only receives particles due to drifts as it is not magnetically connected to the injection region. 
There is an increase in the decay rate after ${\sim} 32$ hours in the highest energy channel.
This faster decay is not present when drifts are not included (see e.g. the left panel of Fig.~\ref{fig:normInt700} for an observer in the equatorial plane). Moreover, the intensities in the $2.70$ MeV channel converge towards the intensities of the well-connected observers located at $0^\circ$ and $-2.25^\circ$ latitude. 

In considering all observers, we see that for the $31.62$ MeV energy channel, the different observers never measure the same intensities, that is, the SEP flood phenomenon is not reproduced. 
When excluding the observer at  $2.25^\circ$ latitude, the flood phenomenon is reproduced during a short interval  (between ${\sim} 4$ and ${\sim} 14$ hours). 
However, in this interval, the reservoir effect is not present in the $2.70$ MeV channel unless also the observer located at $-3.0^\circ$ latitude is excluded.  
We note that all well-connected observers would see the same intensity time-profiles if drifts were not present, since particles are injected uniformly from the injection region.
Thus, these results clearly demonstrate that drifts prevent the SEP flood phenomenon to occur for observers with different magnetic connection to the particle source, at least when no other cross-field transport mechanisms are at work. This applies to the observers at different radial distances (not shown here).

\subsection{The intensity decay rates}\label{sec:decay}
\begin{figure*}
        \centering
        \begin{tabular}{ccc}
        \includegraphics[width=0.31\textwidth]{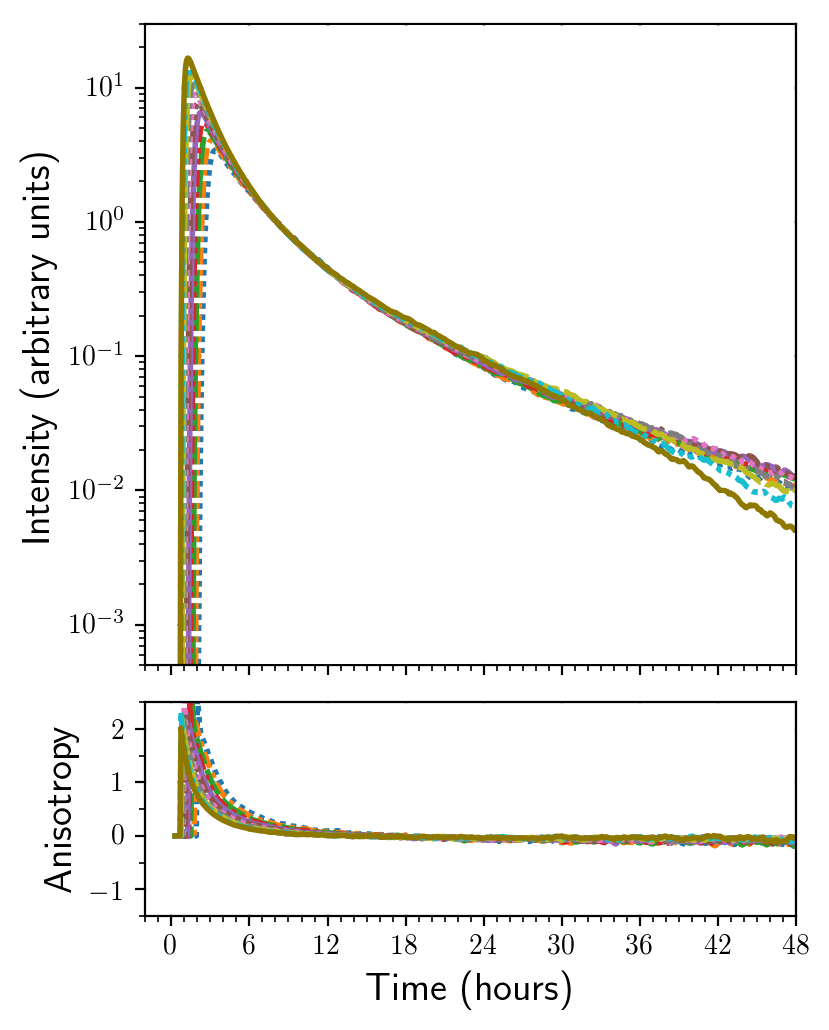} & 
        \includegraphics[width=0.31\textwidth]{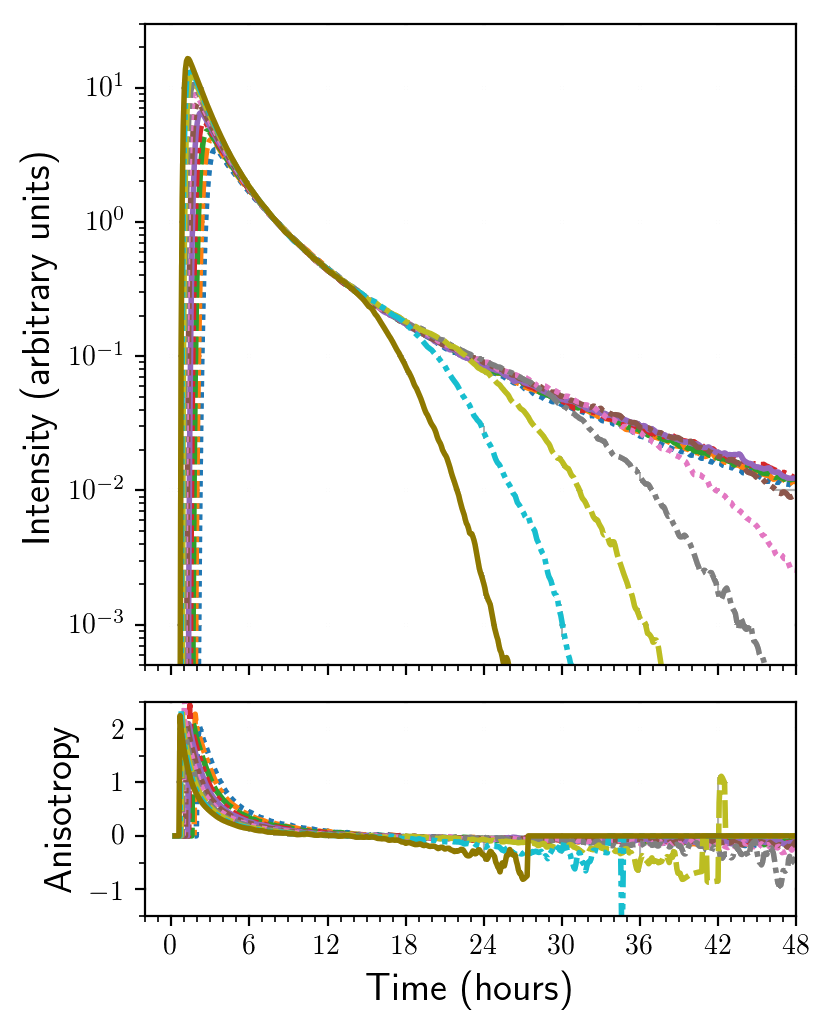} &
        \includegraphics[width=0.31\textwidth]{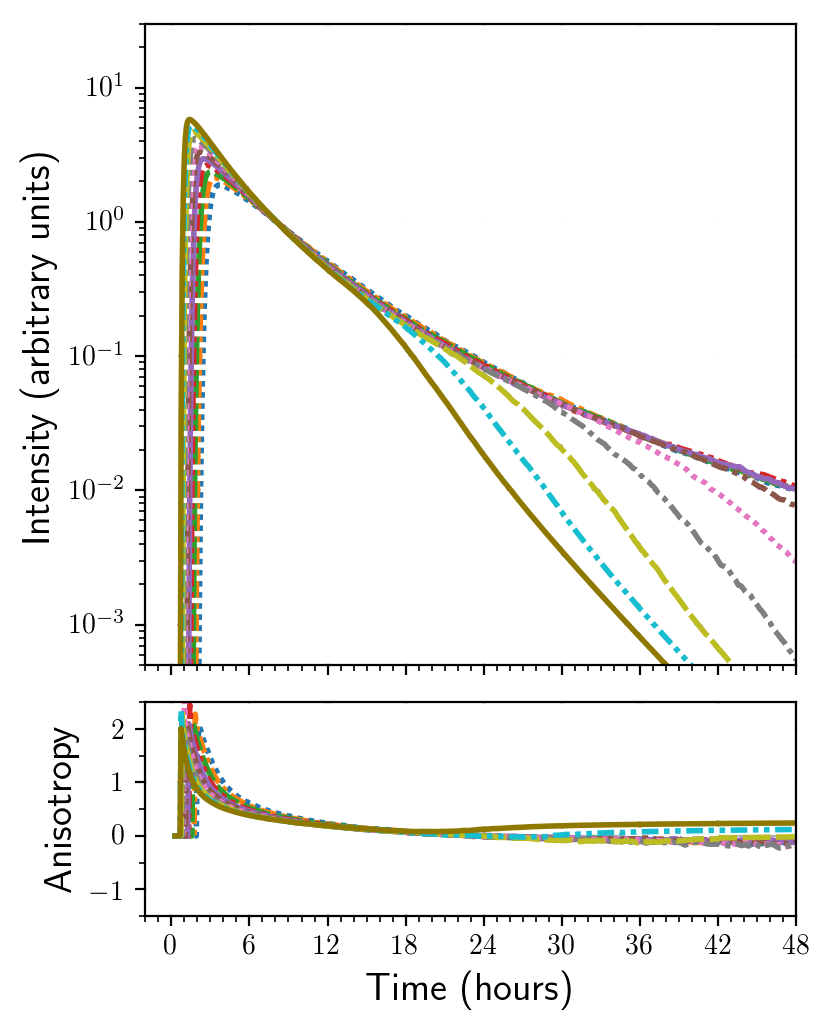} 
    \end{tabular}
    \includegraphics[width=1\textwidth]{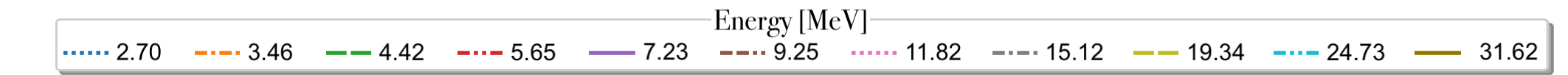}
    \caption{Normalised intensity-time profiles for observers at $0^\circ$ latitude. Left panel is for simulations with no particle drifts, whereas the central panel includes drifts.  In the right panel, particles are injected according to a Reid-Axford profile with a slower decay rate (see text).    }
     \label{fig:normInt700}
\end{figure*}
As noted in the introduction, the decay phase of SEP events is often  characterised by disappearing spatial intensity gradients and by identical decay rates at different energy channels. 
In this section, we compare these decay rates by normalising the intensity time-profiles obtained for an observer located in the solar equatorial plane to their value attained at $t=8$~hours.
As in the previous section, we consider a 700~km~s$^{-1}$ solar wind with $\lambda_\parallel^r = 0.3$ AU. 
The left panel of Fig.~\ref{fig:normInt700} shows the intensity-time profiles obtained when the particle drift $\vec{V}_d$ in Eq.~\eqref{eq:fte_x} is set to zero in the simulation. 
In this case, particle intensity measured in different energy channels decay at a similar rate until nearer to the end of the simulation where the highest energy channel starts decaying faster.
In spite of the  particles having been injected up to 60~MeV, eventually the majority have decelerated to energies below 31.62~MeV and, consequently,  this energy channel becomes depleted of particles. 
The depletion of intensity in the 31.62~MeV channel does not occur within the simulated 48 hours when, instead, particles are injected up to, for example, 90~MeV, (not shown here).
The central panel of Fig.~\ref{fig:normInt700} shows the intensity-time profiles  when the drift term is included in Eq.~\eqref{eq:fte_x}, illustrating that drifts cause a significant increase in the decay rate in the higher energy channels. 
Consequently, it is only approximately ten hours after the prompt phase that a similar decay rate is seen in all energy channels. 

Finally, the right panel of Fig.~\ref{fig:normInt700} shows the intensity-time profiles when injecting particles according to a Reid-Axford profile with a much slower decay rate, that is, with $\beta = 0.2 $~hours  $\tau=8$~hours in Eq.~\ref{eq:RA}. Such a heavy-tailed injection time-profile mimics a prolonged injection of SEPs.
As can be seen, the intensity drop-off in the high-energy channels due to drifts is much more gradual.
However, different energy channels still show different decay rates. 

\subsection{Peak-intensity energy spectra}\label{sec:erg_peak}
\begin{figure}
        \centering
        \includegraphics[width=0.49\textwidth]{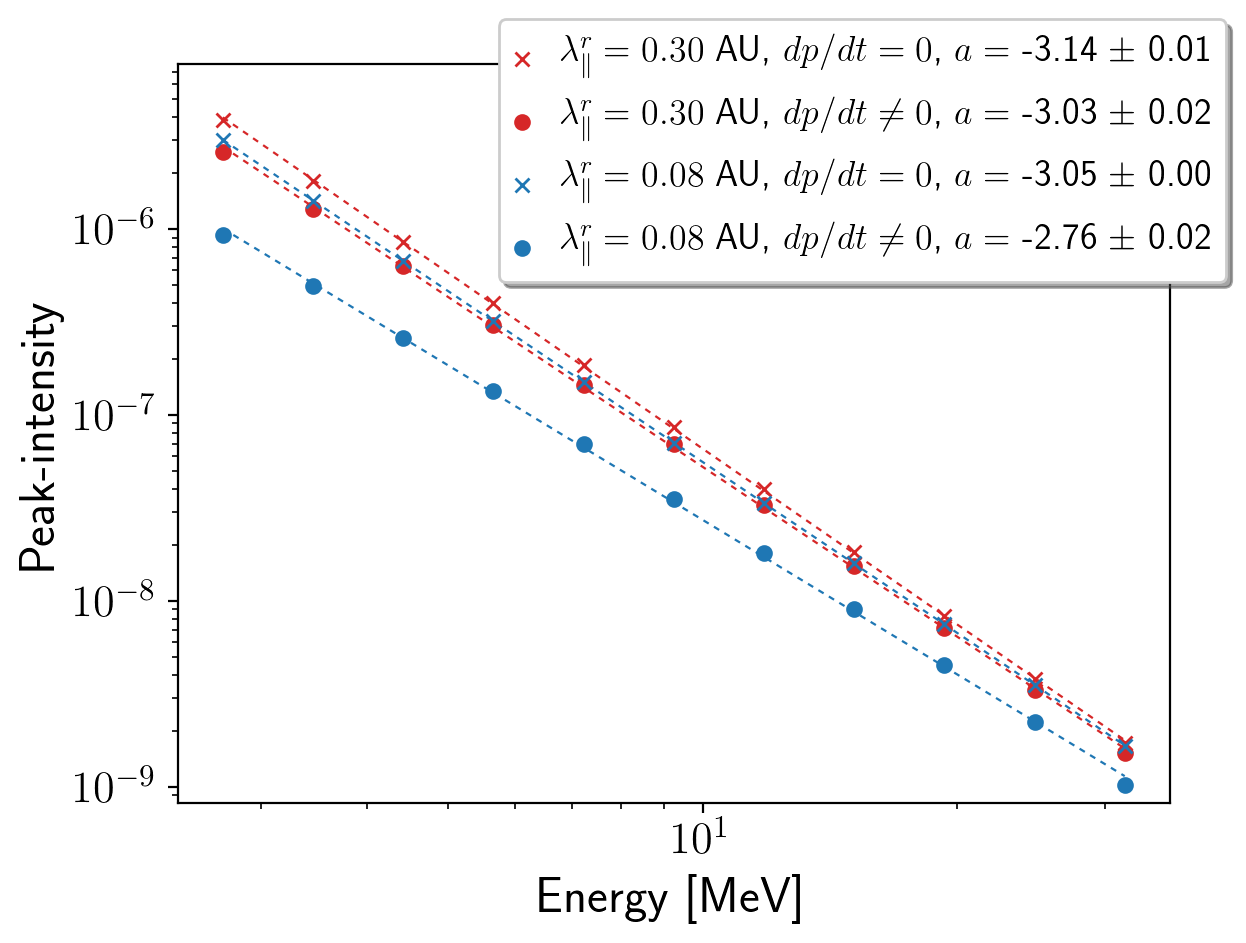} 
    \caption{Peak-intensity spectra for simulations with $\vec{V}_d = 0$ in Eq.\eqref{eq:fte_x} and for observers at 1.0 AU and $0^\circ$ latitude in the 700~km~s$^{-1}$ solar wind.
     Dotted lines are linear fits with the slope $a$ and the standard error given in the inset.}
     \label{fig:central_energy_spectrum}
\end{figure}
\begin{figure*}
        \centering
        \begin{tabular}{cc}
        \includegraphics[width=0.47\textwidth]{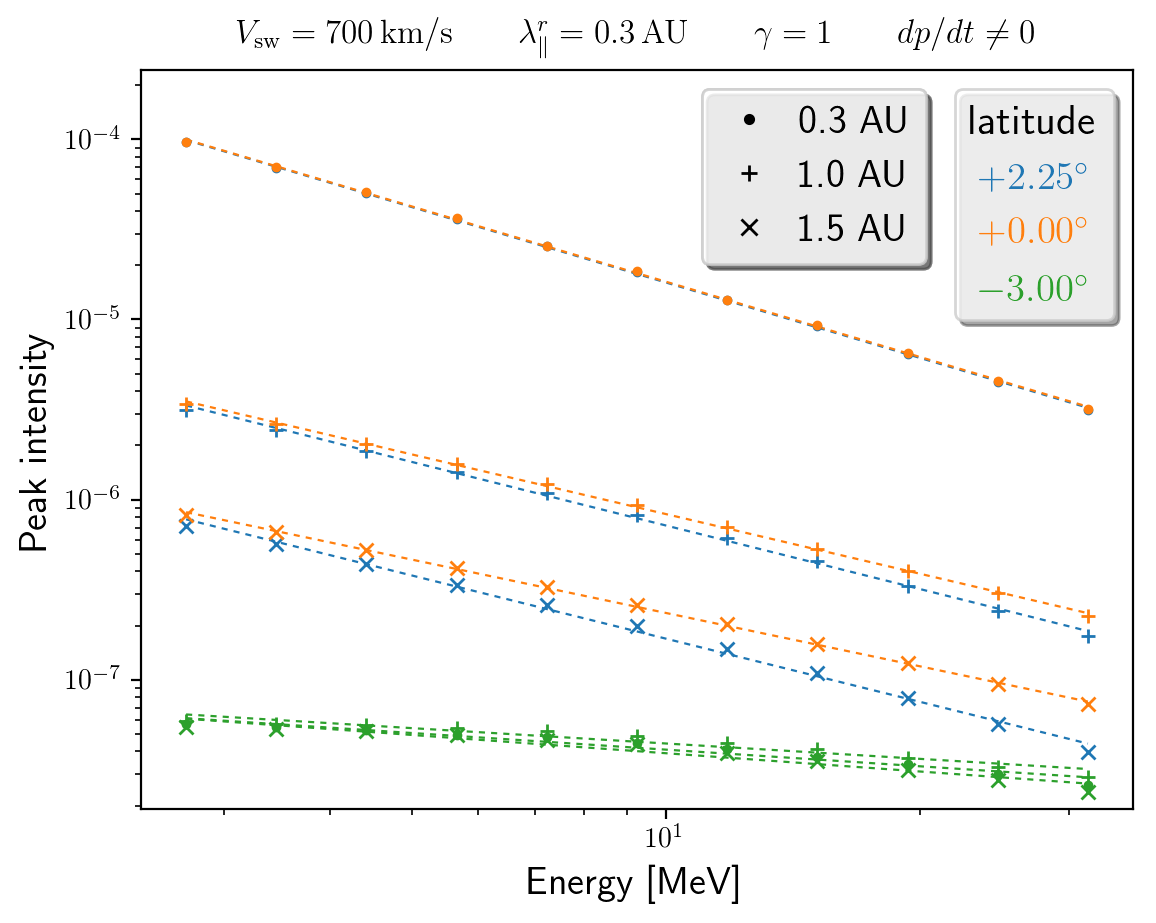} & 
        \includegraphics[width=0.47\textwidth]{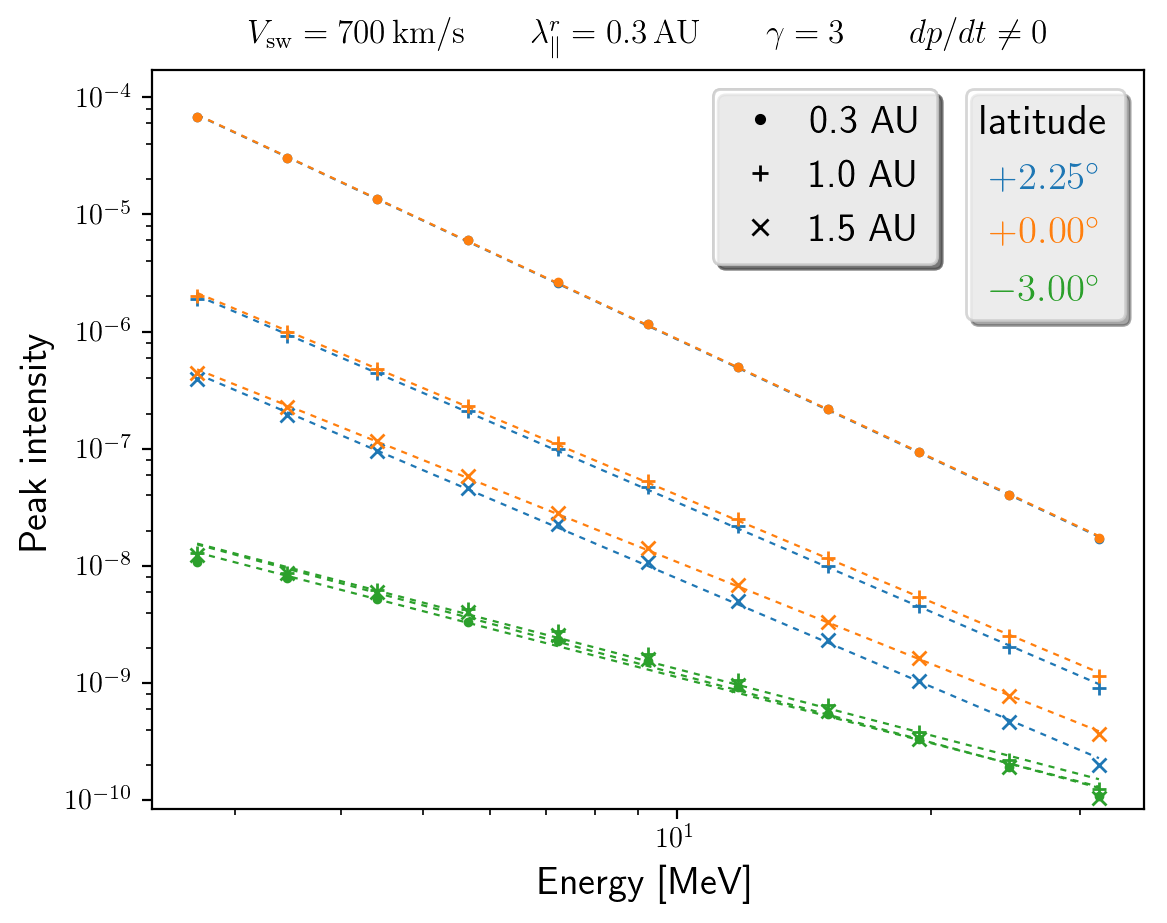} 
    \end{tabular}
    \caption{Peak-intensity spectra for simulations with the effect of drifts included and for observers in a 700~km~s$^{-1}$ solar wind, located at different radial distances and latitudes, as indicated by the legend. Left and right panels are for injection intensities $j\propto E^{-1}$ and $j\propto E^{-3}$ , respectively. 
    The dotted curves are linear fits with the slope $a$ and its standard error given in Table~\ref{table:1}. We note that the orange line covers the blue line for the observers at 0.3 AU (dot symbols). }
     \label{fig:peakEnergySpectrum}
\end{figure*}

In this section, we study the effects of drifts on the energy spectrum measured during the peak-intensity in every energy channel for a 700~km~s$^{-1}$ solar wind. 
The peak-intensity energy spectrum is often used as a proxy for the particle energy spectrum $dN/dE$ near the Sun, where $dN$ denotes the number of particles in an infinitesimal phase-space element\footnote{The differential energy spectrum, $j(E)$ (often denoted as $dJ/dE$), is proportional to $\varv \, dN/dE$. }. 
If particles were to travel scatter-free and without energy changes, the observed peak-intensity energy spectrum would correspond to the injected spectrum. In the present simulation setup, we would therefore obtain a peak intensity spectrum proportional to $E^{-3.5}$, since particles are injected according to $j \propto E^{-3}$ (or $dN/dE \propto E^{-3.5}$; see Section 2).  

However, adiabatic deceleration and the (energy dependent) particle scattering will harden the spectrum. Figure~\ref{fig:central_energy_spectrum} shows the energy spectra for four simulations without particle drifts, both with and without particle deceleration and for two different radial mean free paths. 
Denoting the slope of the linear (in log-log space) fit to the data in the figure by $a$, we see that all energy spectra are harder than the injected spectrum, that is, the absolute value of the slope is always smaller than 3.5.
The energy dependence of the assumed diffusion coefficient $D_{\mu\mu}$ (see also Eq.~(8) in \cite{wijsen19a}) implies that low-energy particles have a smaller mean free path than high-energy particles, therefore lowering the attained peak-intensity more for the former compared to the latter \citep{lario07}. As a result,  the energy spectrum is hardened.
The diffusion of the peak-intensities due to scattering also explains why in Fig.~\ref{fig:central_energy_spectrum}, the peak-intensities are lower for $\lambda_\parallel^r = 0.08$ AU  than for $\lambda_\parallel^r = 0.3$ AU. 
Moreover, the smaller the mean free path, the more pronounced the effect of the energy dependence of the mean free path on the energy spectrum as it takes a longer time for the particles to reach the observer. 
In addition, the smaller mean free path for low-energy particles will keep these particles closer to the Sun for a prolonged time, where the adiabatic deceleration is largest and, hence, further hardening the energy spectrum. 

Apart from scattering and adiabatic deceleration, particle drifts can modify the observed peak-intensity energy spectrum, as drifts are energy dependent. 
Figure~\ref{fig:peakEnergySpectrum} shows the peak-intensity energy spectra for observers that are, as before, located at different radial distances along the same IMF line. 
This is done for three different IMF lines, where each IMF line's foot-point has a different latitude.
The left and right panels of Fig.~\ref{fig:peakEnergySpectrum} shows results for the injection intensity $j(E) \propto E^{-1}$ and $j(E) \propto E^{-3}$, respectively. The dotted lines are linear fits to the energy spectra, with the estimated slope $a$ and the standard error given in  Table~\ref{table:1}. 

For observers in the solar equatorial plane (orange curves) the energy spectra can be seen to harden with radial distance.
In contrast, for the observer located at $2.25^\circ$ latitude (blue curves), the energy spectra remain constant with radial distance. 
This is because by the time the particles reach 1.5~AU, some high-energy particles have already drifted significantly southward, counteracting the spectral hardening that would be otherwise observed.  
Consequently, an IMF line with foot-point closer to the northern edge of the injection region would show an energy spectrum that softens with radial distance.
 At small radial distances,  the observers located at 0.3~AU with latitudes $2.25^\circ$ and $0^\circ$ measure identical peak-intensities. This is because these peak-intensities are obtained early in the simulation, before the effect of drifts have become notable.

The green curves of Fig.~\ref{fig:peakEnergySpectrum} show that observers at latitude $-3^\circ$ have a much harder spectrum than the others. 
This is due to the proportionality of the drifts to the particle energy. 
In the left panel, depicting the case of $\gamma = 1$ in Eq.~\eqref{eq:RA}, we expect, therefore, to see a peak-intensity energy spectrum  $ E^{-0.5}$. 
However, as noted before, particle scattering and adiabatic deceleration will harden the spectrum even more, explaining the $a < 0.5$ slope that is obtained. Analogously for  case of $\gamma = 3$, we expect to see a peak-intensity energy spectrum harder than $E^{-2.5}$, which is indeed the case.
Finally, it is interesting to note that the energy spectrum of observers along the IMF line with latitude $-3^\circ$ is independent of radial distance. This is because the peak-intensity occurs during the decay phase of the SEP event when the SEP flood phenomenon is established along IMF lines (see Section~\ref{sec:Imf_obs}). 

\begin{table}
\caption{Linear Regression variables of Fig.~\ref{fig:peakEnergySpectrum}.}              
\label{table:1}      
\centering                            
\begin{tabular}{c c c c c}          
\hline\hline    
$\gamma$ & lat [$^\circ$] & $r$ [AU] & slope $a$ & standard error  \\   
\hline                                   
1  & \phantom{+}2.25 & 0.3 & $-1.39$ & 0.01 \\     
   & \phantom{+}2.25 & 1.0 & $-1.17$ & 0.01 \\
   & \phantom{+}2.25 & 1.5 & $-1.17$ & 0.01 \\
   & \phantom{+}0.00 & 0.3 & $-1.39$ & 0.01 \\
   & \phantom{+}0.00 & 1.0 & $-1.10$ & 0.01\\
   & \phantom{+}0.00 & 1.5 & $-0.98$ & 0.01 \\
   & $-3.00$ & 0.3 & $-0.33$ & 0.03  \\
   & $-3.00$ & 1.0 & $-0.29$ & 0.02 \\
   & $-3.00$ & 1.5 & $-0.34$ & 0.03 \\
  \hline
 3 & \phantom{+}2.25 & 0.3 & $-3.37$ & 0.01 \\     
   & \phantom{+}2.25 & 1.0 & $-3.10$ & 0.02 \\
   & \phantom{+}2.25 & 1.5 & $-3.07$ & 0.03 \\
   & \phantom{+}0.00 & 0.3 & $-3.37$ & 0.01 \\
   & \phantom{+}0.00 & 1.0 & $-3.03$ & 0.02 \\
   & \phantom{+}0.00 & 1.5 & $-2.89$ & 0.02\\
   & $-3.00$ & 0.3 & $-1.94$ & 0.05 \\
   & $-3.00$ & 1.0 & $-1.89$ & 0.05 \\
   & $-3.00$ & 1.5 & $-1.94$ & 0.04 \\
\hline                                             
\end{tabular}
\end{table}

\subsection{Energy spectra of fluence}\label{sec:erg_fluence}
\begin{figure*}
        \centering
        \begin{tabular}{ccc}
        \includegraphics[width=0.31\textwidth]{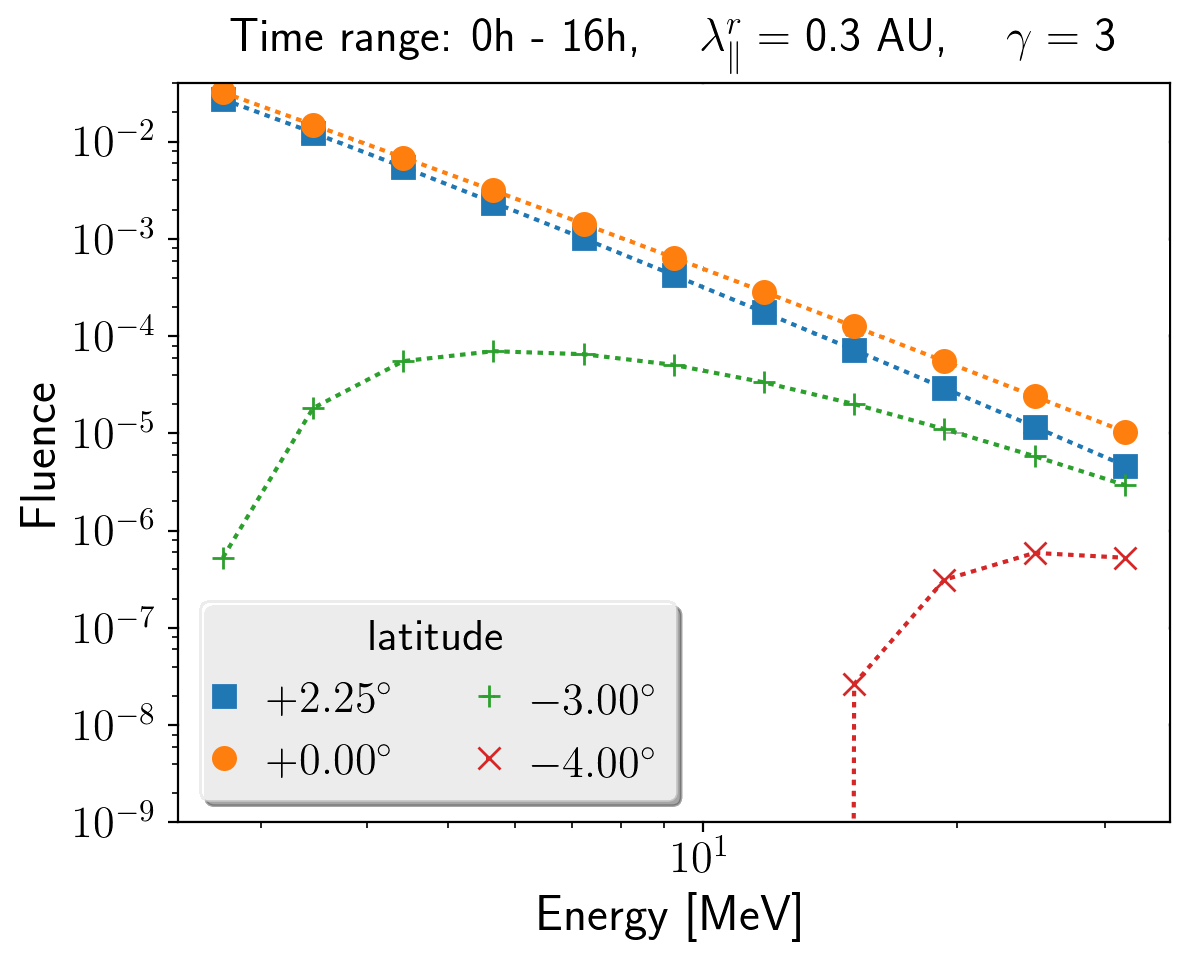} & 
        \includegraphics[width=0.31\textwidth]{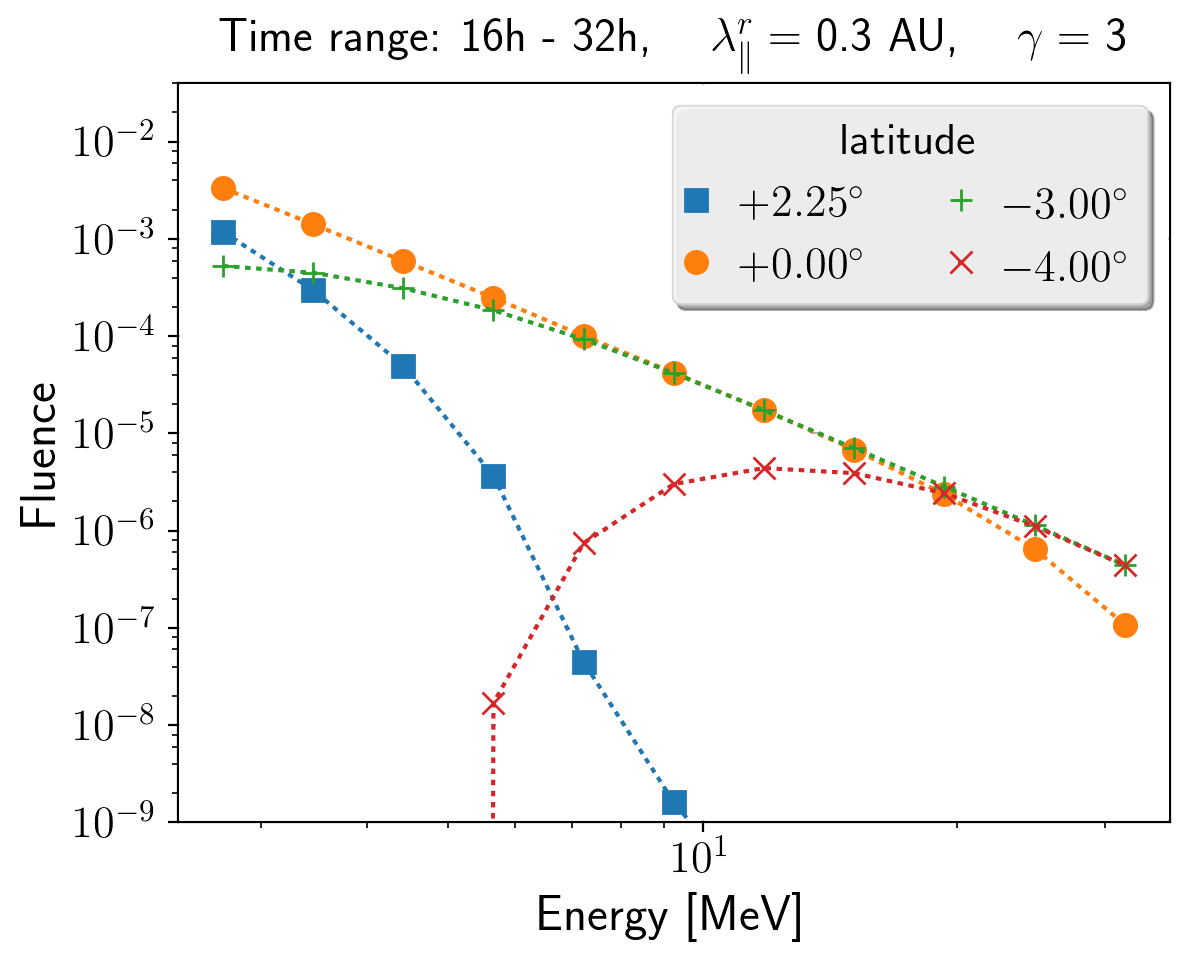} &
        \includegraphics[width=0.31\textwidth]{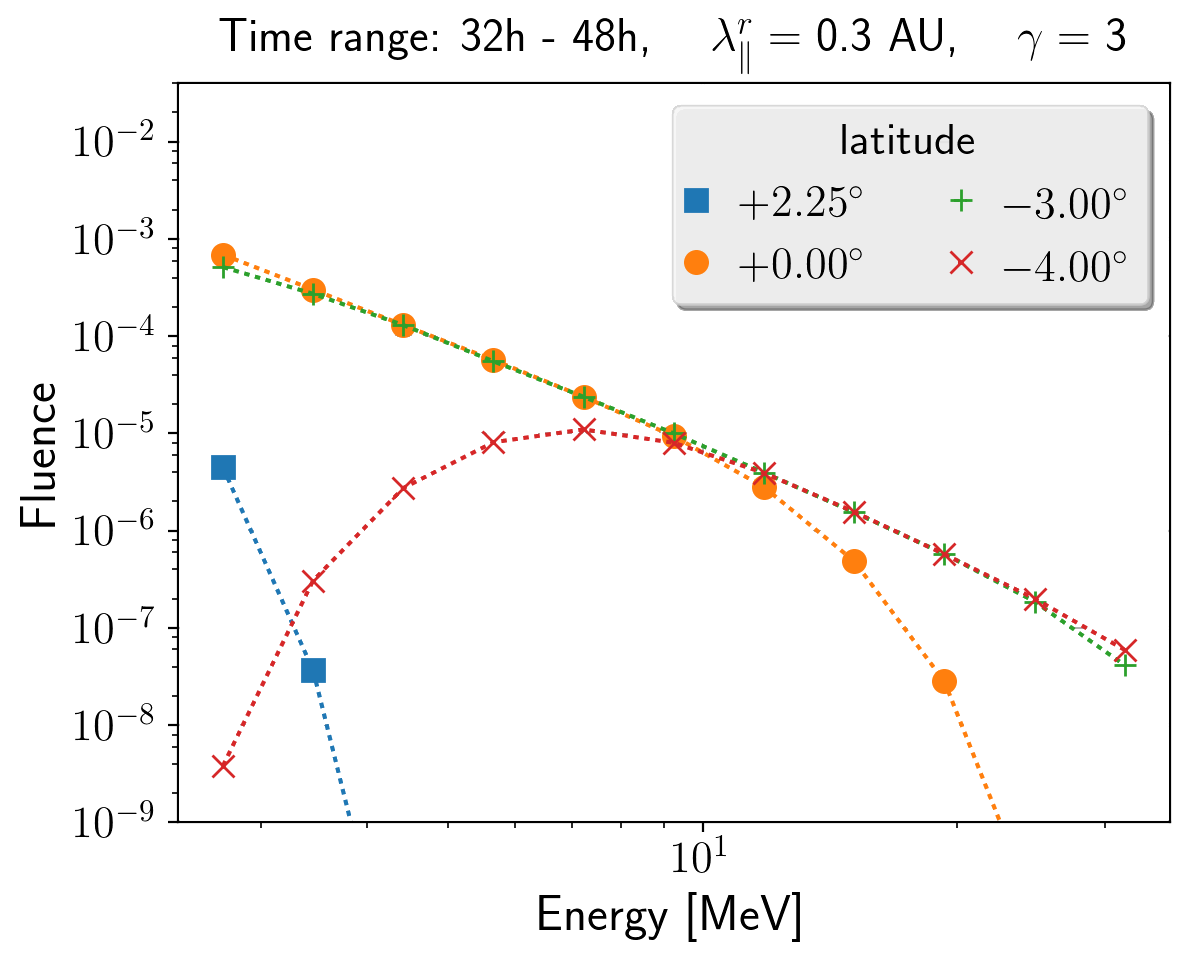}\\ 
    \end{tabular}
    \caption{ Fluence energy spectra for three different time intervals, as indicated above the figures. The energy spectra are for observers in a 700~km~s$^{-1}$ Parker wind located at different latitudes at 1 AU.}
     \label{fig:fluence700}
\end{figure*}  
We now examine the energy spectrum obtained from computing the particle fluence for observers located at different latitudes at 1 AU.
As for the case with the energy spectrum of the peak-intensities, the fluence energy spectrum can be used as a proxy for the energy spectrum $dN/dE$ near the Sun, especially when the particle mean free path is large.

Since the particle drifts introduce changes on the timescale of hours, we divide the 48 simulated hours in three different time intervals, each 16~hours long.
Figure~\ref{fig:fluence700} shows the energy spectra for four different observers. 
In the left panel, we see that during the first 16-hour interval, drifts soften the energy spectrum for the observer at $2.25^\circ$ latitude with respect to  the observer at  $0^\circ$. 
For the observer at $-3^\circ$ latitude we see an energy spectrum with positive slope at low energies, again due to the energy dependence of the drifts  (drifted low-energy protons have just started to reach the 1.0~AU observer within this period; see the left panel of Fig.~\ref{fig:3obsImf700}). 
The central panel of Fig.~\ref{fig:fluence700} shows how a very soft energy spectrum is observed at $2.25^\circ$ latitude. 
In contrast, the energy spectra of the observers at $-3^\circ$ and $-4^\circ$ latitude converge towards the one of the observer in the solar equatorial plane, except at the high-energy end, where the equatorial observer starts seeing a softening of its energy spectra due to the particle drifts.
This becomes even more pronounced in the last 16 hours of the simulation, during which the energy spectrum of the $0^\circ$ observer is formed by a power law and an exponential rollover at $\sim 10$ MeV (see central panel of Fig.~\ref{fig:normInt700}).
 
\subsection{The effect of cross-field diffusion}\label{sec:cfd}
\begin{figure*}
        \centering
        \begin{tabular}{ccc}
        \includegraphics[width=0.31\textwidth]{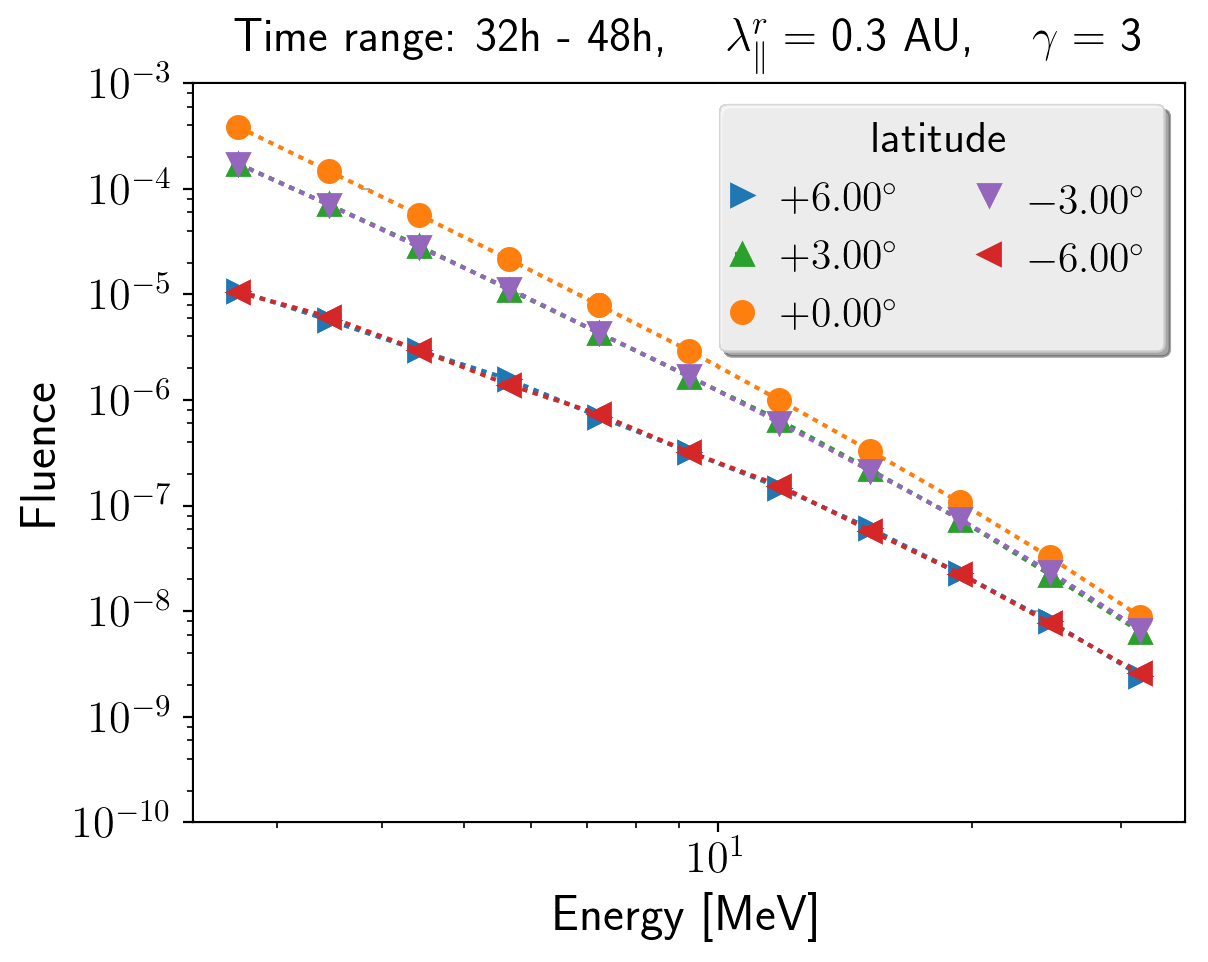} &
\includegraphics[width=0.31\textwidth]{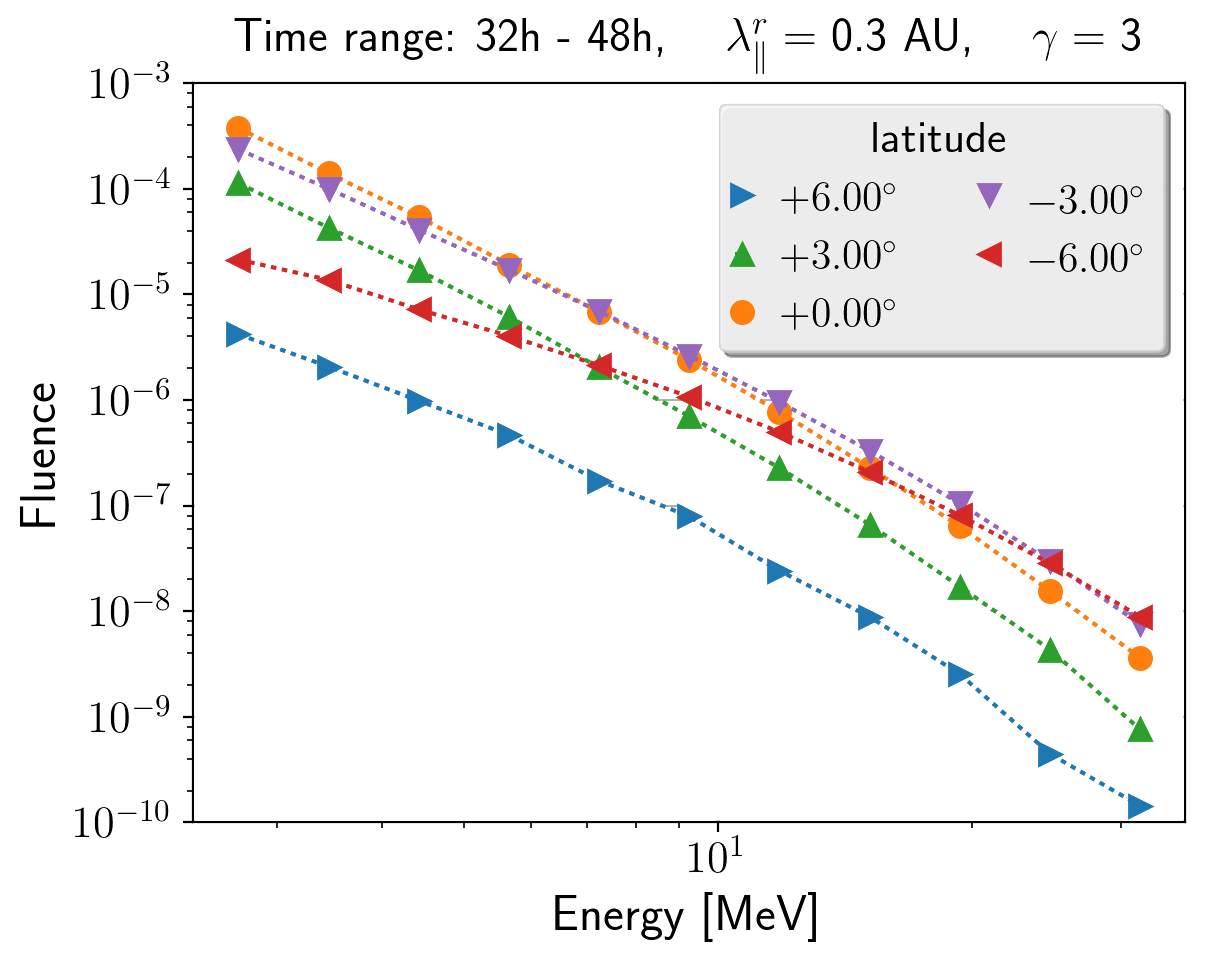}&
\includegraphics[width=0.31\textwidth]{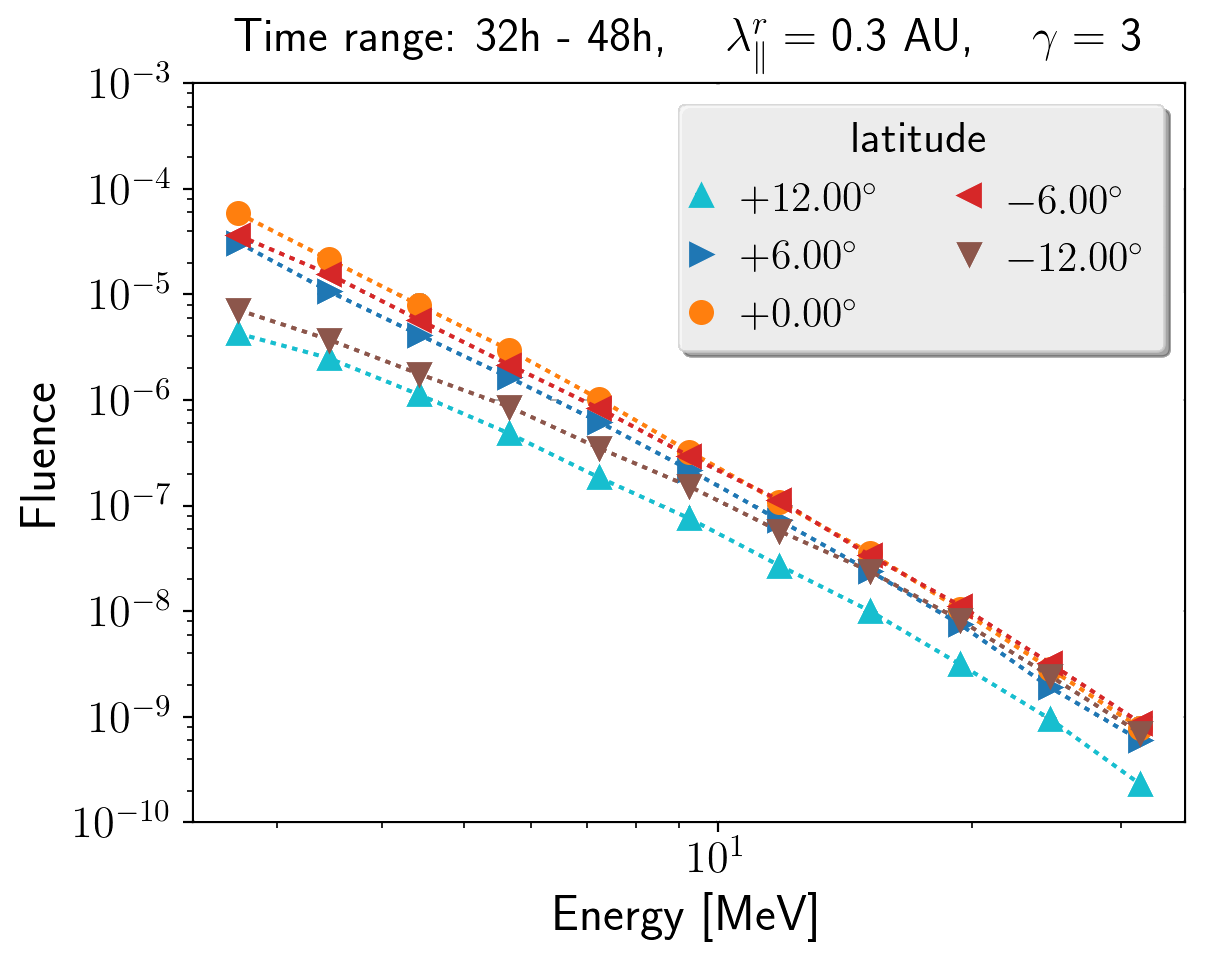}
    \end{tabular}
    \caption{Fluence energy spectra for simulations with cross-field diffusion, measured between 32h and 48h. Left panel and central panel are for $\alpha = 10^{-4}$, whereas the right panel is for $\alpha = 10^{-3}$. Left panel is for simulations without particle drifts, whereas the central and right panel include drifts.  }
     \label{fig:K_fluence700}
\end{figure*}

In the previous sections, we illustrate how drifts  alter the modelled intensity-time profiles and energy spectra. 
In particular, it is shown that drifts can pose a challenge for the formation of the SEP flood phenomenon, if no other cross-field transport mechanisms are available. 
The SEP flood or reservoir effect is, however, attributed as an effect of spatial diffusion \citep[e.g.][]{wang15}.
In this section, we aim to show how cross-field diffusion might alter the results of the previous sections. In order to  do so, we perform simulations using an axis-symmetric cross-field diffusion process, with the following diffusion coefficient \citep{droge10, wijsen19c}
\begin{eqnarray}\label{eq:cross_field}
\bm{\kappa}_\perp &=&\alpha \frac{\pi }{12}{ {\varv}\lambda_\parallel}\frac{B_{0}}{B},
\end{eqnarray}
where $B_{0}$ is the magnetic field strength at 1 AU, $\lambda_\parallel = \lambda_\parallel^r/ b_r^2$ is the parallel mean free path, and the parameter $\alpha$ determines the strength of the cross-field diffusion.  
The simple functional form of $\kappa_\perp$ allows us to easily track the effects of cross-field diffusion in our simulations. 
We leave it to future work to study the effects of more advanced cross-field diffusion theories on the pitch-angle dependent drifts.

The left and central panels of Fig.~\ref{fig:K_fluence700} show the fluence energy spectra for different observers, measured between 32h and 48h and with a cross-field diffusion parameter $\alpha = 10^{-4}$.   
The left panel of the figure corresponds to simulations without particle drifts, whereas the central panel includes the effects of particle drifts. 
A comparison of both panels illustrates that significant differences exist between the two simulations.
The symmetry with respect to the solar equatorial plane is broken when drifts are present, with the observers in the southern hemisphere measuring a harder energy spectrum than the corresponding observers in the northern hemisphere. 
Since different observers measure a different energy spectrum, the SEP flood phenomenon is not reproduced.  
The slopes of the energy spectra of the different observers match more closely when no particle drifts are included, although they are not identical. This is because the cross-field diffusion model we use contains an energy-dependence,  which counteracts the establishment of the reservoir phenomenon. 
In contrast, in other approaches,  such as the meandering field line model of \cite{laitinen13}, the cross-field diffusion is independent of the particle's energy and would, hence, spread the particles throughout the heliosphere without changing the energy spectrum.

The right panel of Fig.~\ref{fig:K_fluence700} shows the same as the central panel, yet for a cross-field diffusion process that is ten times stronger ($\alpha = 10^{-3}$). 
As before, we see that the particle drifts can break the symmetry with respect to the solar equatorial plane. However, the difference between the observers at $\pm 6^\circ$  is considerably smaller than in  the central panel, that is, cross-field diffusion mitigates the effect of the drifts.
At larger latitudes and high energies, we still see a non-negligible difference between the northern and the corresponding southern observer. 

\subsection{A magnetic bottle}\label{sec:cir}
\begin{figure}
        \centering
        \includegraphics[width=0.47\textwidth]{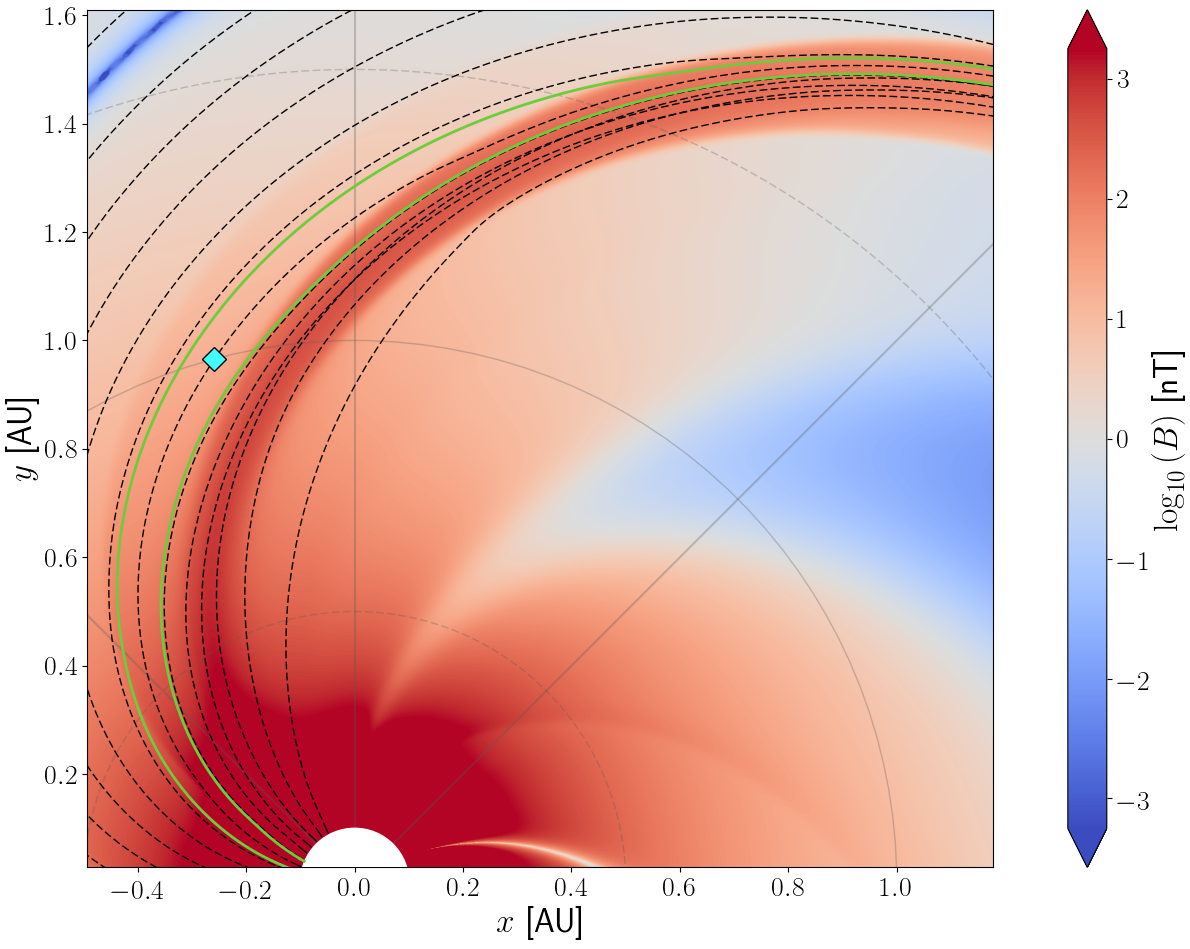} 
    \caption{Magnetic field strength in the solar equatorial plane. Dashed lines represent some IMF lines with equally spaced foot-points. Green lines are the IMF lines at the edges of the particle injection region. The cyan diamond is the observer discussed in the text. }
     \label{fig:cir}
\end{figure}
\begin{figure*}
        \centering
        \begin{tabular}{ccc}
        \includegraphics[width=0.31\textwidth]{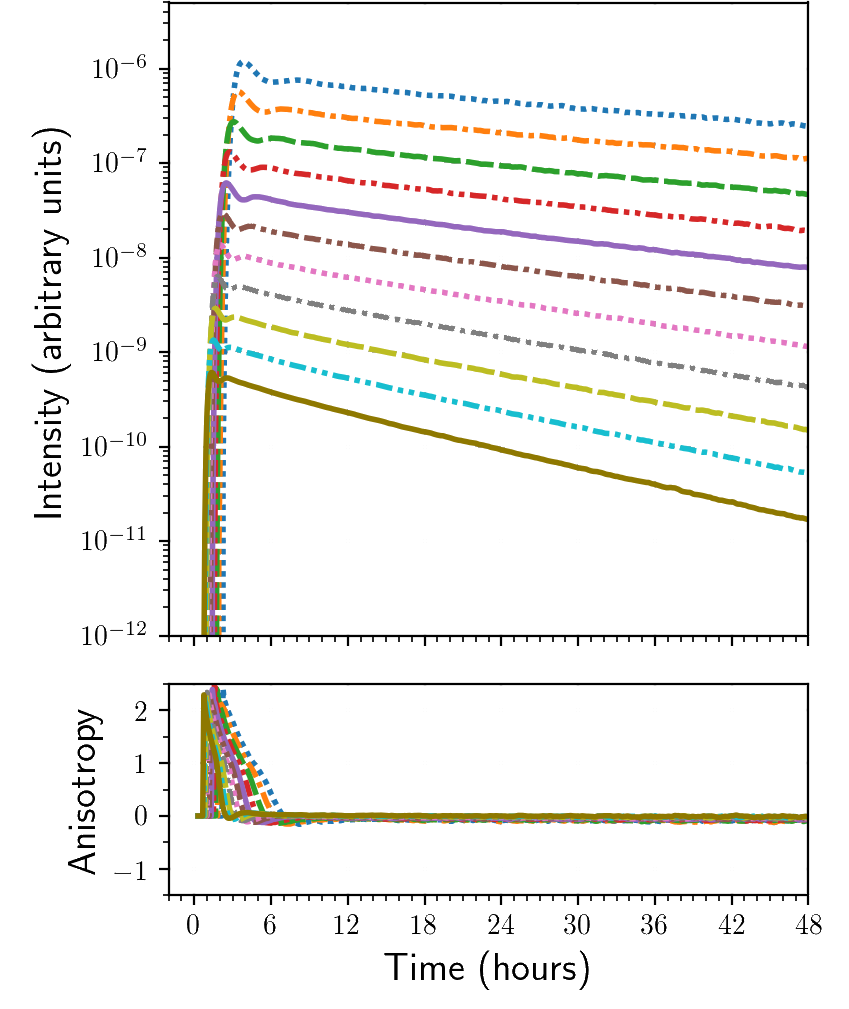} & 
        \includegraphics[width=0.31\textwidth]{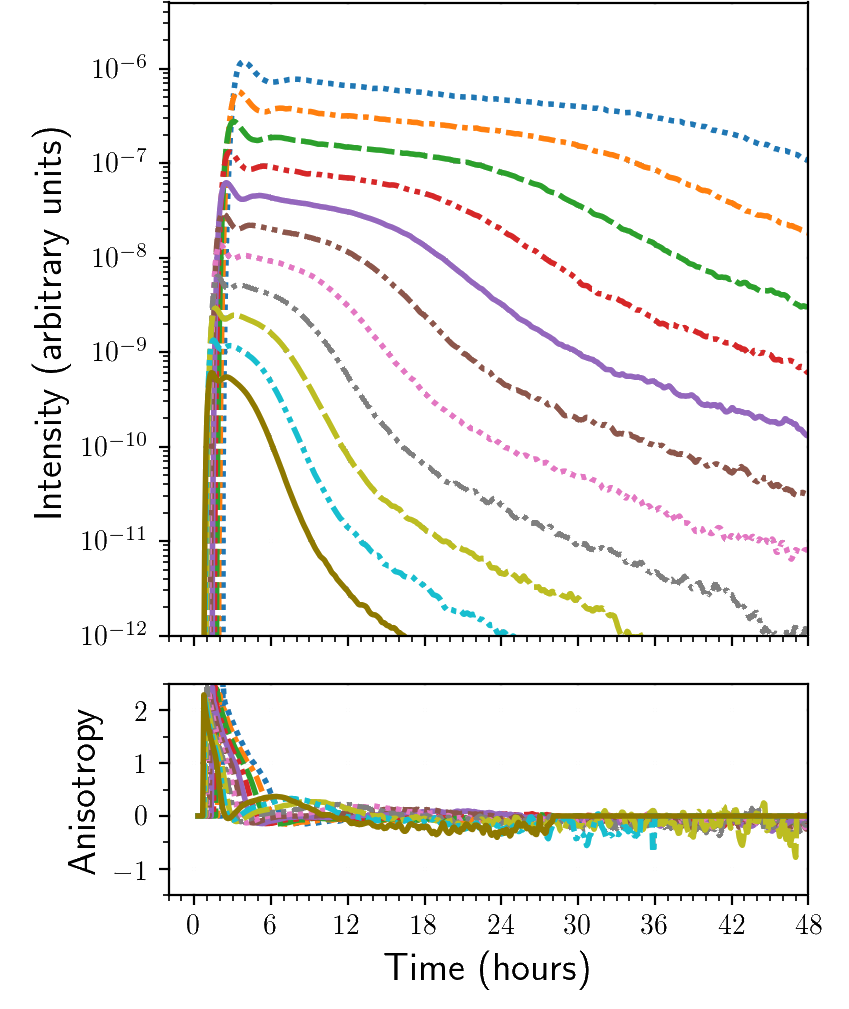} &
        \includegraphics[width=0.31\textwidth]{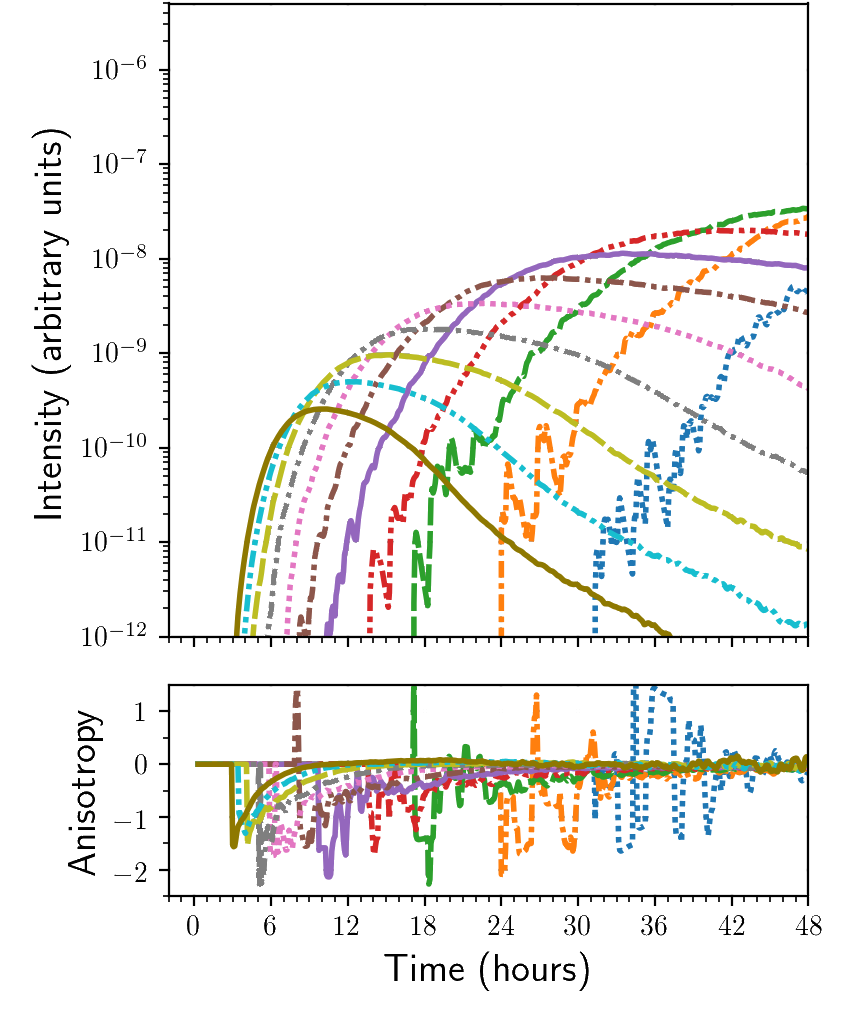} 
    \end{tabular}
    \includegraphics[width=1\textwidth]{legend.png}
    \caption{Left and central panels are the intensity-time profiles for the observer indicated by the cyan diamond in Fig.~\ref{fig:cir} for simulations without and with particle drifts, respectively. The right panel is for an observer at the same longitude as the other panels yet at a latitude of $-6.5^\circ$ and with the effect of particle drifts included. }
     \label{fig:intCir}
\end{figure*}
The reservoir phenomenon has usually been attributed to the expansion of a magnetic bottle, typically in association with CMEs \citep{reames96}. 
Another solar wind configuration that also contains a magnetic bottle is a CIR.
In this case, one end of the bottle is formed by the converging IMF lines near the Sun whereas the other end is formed at the CIR, where the plasma and IMF are compressed.
In \cite{wijsen19b, wijsen19a} we modelled a CIR by using the 3D MHD model EUHFORIA. 
This was done by prescribing a slow solar wind of 330~km~s$^{-1}$ everywhere at the inner boundary, with the exception of a circular 660~km~s$^{-1}$ fast solar wind region with a $30^\circ$ diameter and a midpoint located at $5^\circ$ latitude (see e.g. Fig.~1 of \citeauthor{wijsen19a}~\citeyear{wijsen19a}). 
Between the slow and fast wind, a transition region exists in which the solar wind speed increases linearly. 
This transition region eventually evolves at larger radial distances into a corotating interaction region bounded by two MHD shock waves. For more details, we refer to \cite{wijsen19b} and \cite{wijsen19a}.

Figure~\ref{fig:cir} shows the logarithm of the magnetic field strength in the solar equatorial plane in the MHD simulation, as well as some magnetic field lines. From the inner boundary, we inject particles in a region located in the slow solar wind in front of the fast wind stream. 
The magnetic field lines bounding the injection region are denoted in green and, as before, the injection region is centred on the solar equatorial plane with a latitudinal width of $5^\circ$.
The green IMF lines in Fig.~\ref{fig:cir} highlight the magnetic bottle structure that has formed. In addition, it is shown in \cite{wijsen19a} that the strong and sudden magnetic field enhancement at the forward shock or compression wave of the CIR is an effective particle mirror. 
Particles are injected as described in Section~\ref{sec:modelling}, with $\lambda_\parallel^r = 0.3 $ AU and $\kappa_\perp = 0$. Moreover, we do both simulations with and without the inclusion of particle drifts in the spatial term of the FTE. 

The left and central panel of Fig.~\ref{fig:intCir} show the intensity-time profiles for the corotating observer indicated by a cyan diamond in Fig.~\ref{fig:intCir}, who is located in the solar equatorial plane. The left panel shows the results obtained for simulations when particle drifts are neglected in Eq.~\eqref{eq:fte_x}, whereas the central panel includes those particle drift effect. 
The time-intensity profiles in the left panel decay very slowly with time. For example, over a range of $\sim 48$ hours the intensities only drop by a factor of $\sim 3$ for the lowest energy channel, which is much less than in a Parker spiral, where there is typically a decrease of several orders of magnitude (see e.g. Fig.~\ref{fig:3obsImf300}). As we already alluded to previously, this is the effect of the magnetic bottle, trapping the particles between the Sun and the CIR. 

In the central panel of  Fig.~\ref{fig:intCir}, it can be seen that the time-intensity profiles are strongly altered when the effect of particle drifts are included. 
At the forward shock wave of the CIR, there is a strong magnetic gradient pointing  sunward, which enhances the  southward  gradient drift motion of the particles.
As a result, all energy channels are affected by the drifts.  This is in sharp contrast with the  300~km~s$^{-1}$ Parker solar wind case, where drifts had no effects on the intensity decay rates for a well-connected observer (see left panel of Fig.~\ref{fig:3obsImf300}), and with the 700~km~s$^{-1}$ solar wind, where only  the high energy channels were affected by drifts (not shown). 
Moreover, the central panel of  Fig.~\ref{fig:cir} shows that different energy channels have different decay rates, with the higher energy channels decaying faster.  No reservoir phenomenon is thus established, despite the presence of a magnetic bottle. 

The right panel of Fig.~\ref{fig:cir} displays the intensities measured by an observer located at $-6.5^\circ$ in latitude. 
In the Parker spiral, this observer did not see any significant intensities, yet here we see that the observer detects  intensities comparable and even higher than the intensities of a well-connected observer in the decay phase. The intensity in the $31.62$ MeV channel is only a factor of $\sim 3$ lower than the peak intensity of that channel for the well-connected observer. 
From the anisotropies, we see that most particles are streaming sunward, which is a result of the significant drifts and the  mirroring that these particles undergo when reaching the  forward shock wave bounding the CIR. 

\begin{figure}
    \centering
    \includegraphics[scale = 0.8]{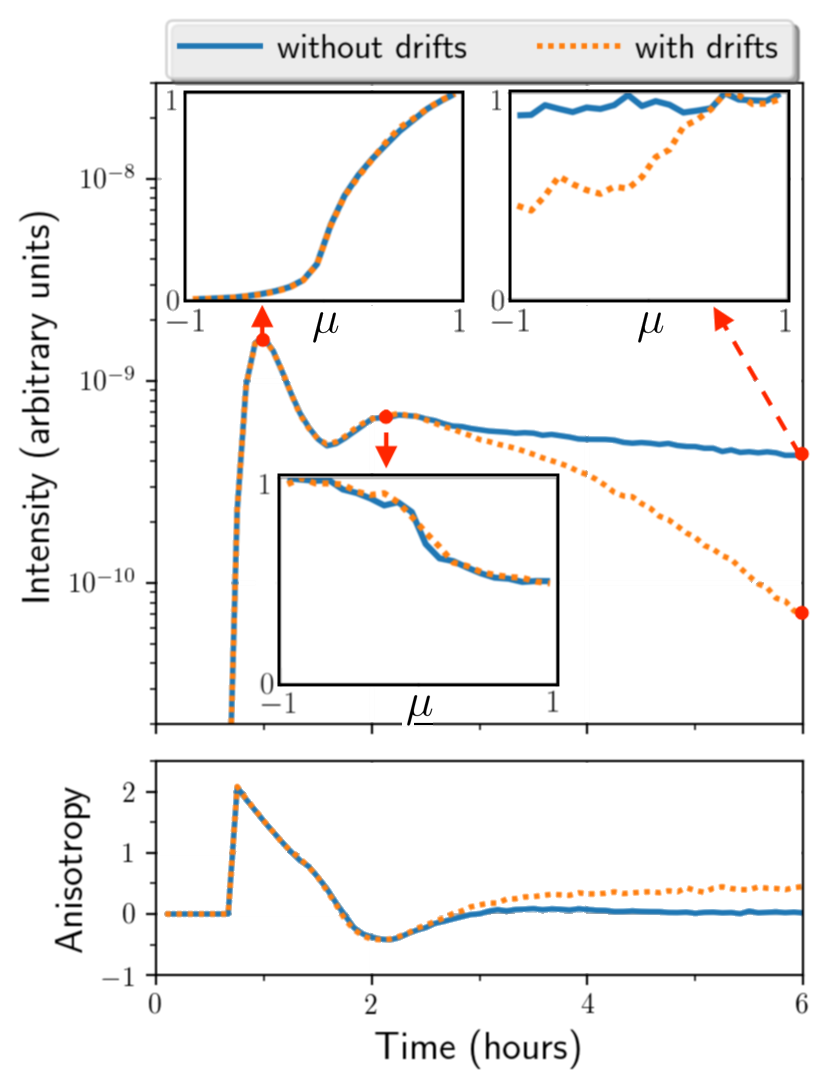}
    \caption{$31.62$~MeV intensities and anisotropies for the observer indicated by the cyan diamond in Fig.~\ref{fig:cir}. 
    The three insets show the normalised pitch angle distributions, corresponding to the snapshots indicated by the red dots. Solid and dotted lines are for simulations without and with particle drifts.}
    \label{fig:pads}
\end{figure}

Figure~\ref{fig:pads} shows the $31.62$ MeV intensities and anisotropies during the first 6 hours of the simulation, when particles are instead injected as a delta function in time. 
The inset plots show three snapshots of the normalised pitch-angle distributions (PADs).
The first snapshot corresponds to the peak phase and shows a clear anti-sunward streaming particle distribution (i.e. a PAD with a positive slope). 
However, after $\sim 2$ hours, a second maximum in the intensity profiles appears. 
The corresponding PAD has a negative slope, indicating sunward streaming protons. These are particles that have been mirrored at the forward shock of the CIR. 
Afterwards, the simulation without drifts slowly decreases in intensity and becomes more isotropic (a vanishing anisotropy and a horizontal PAD). In contrast, the simulation that includes drifts shows a rapidly decreasing intensity profile with positive anisotropies and a PAD of positive slope, indicating anti-sunward streaming protons like at the peak phase of the event. This is because the particles not only undergo mirroring at the forward shock, but also strong southwards drifts, driving the particles to lower latitudes.

\section{Summary and discussion}\label{sec:summary}
In this work, we investigate how drifts may affect the intensity-time profiles of SEP events for protons with energies between 2.39 and 35.76~MeV. Since drifts are a relatively slow process compared to the streaming of the particles along the IMF lines, their effect is mainly manifested during the decay phase of SEP events.

When the effect of drifts is neglected in the spatial part of the FTE, the modelled intensity-time profiles for observers located at different radial distances along an IMF line typically show the same intensity fall-off after the prompt phase of the particle event. 
Moreover, the decay rate is similar for all energy channels. 
Both of these properties are in accordance with the SEP flood or reservoir phenomenon and are expected in a diffusion-dominated system.  
When magnetic gradient and curvature drifts are modelled in a Parker spiral configuration of positive polarity, particles experience a southwards drift. As a result, an observer magnetically connected close to the northern edge of the particle injection region will see a prompt drop in the intensities.
The proportionality of the drifts to energy leads to an intensity drop that occurs earlier for the more energetic particles. 
As a result, the intensity drop is different from the drop observed when a stationary observer leaves the particle streaming zone due to corotation (see e.g. \cite{droge10, wijsen19a}). This latter intensity drop is more abrupt and occurs quasi-simultaneously for all energy channels, unless there is energy-dependent cross-field diffusion or significant particle drifts in the azimuthal direction. 

Observers that have a magnetic connection south of the particle injection region detect the particle onset at later times when compared to a well-connected observer. 
For these southern observers, the particle onset occurs during the decay phase of the SEP event seen by the well-connected observer and, hence,  observers located along the same IMF line see similar intensities and onset times, irrespective of their radial distance from the Sun. In addition, as a consequence of the energy dependence of the drifts, the difference in the arrival times among particles of different energies is larger as compared to a well-connected event. 

The effect of drifts on the different energy channels is enhanced by adiabatic deceleration since high-energy particles may drift significantly before decelerating to lower-energy channels, giving the false impression that the low-energy particles are drifting considerably.
Such an impression is aided by assuming strong scattering conditions since a small mean free path keeps particles for a prolonged amount of time at smaller radial distances, where the adiabatic deceleration is strongest.
This is especially true for a fast solar wind as the adiabatic deceleration scales with the solar wind velocity in a Parker solar wind.

Observers located south of the particle injection region measure a very hard energy spectrum when looking at the peak intensities. 
In contrast, observers magnetically connected to the northern edge of the particle injection region measure a much softer energy spectrum. 
A similar trend is observed when studying the energy spectra derived from particle fluences. 
An observer at positive latitude sees a very hard energy spectra during the decay phase of the SEP event. 
In contrast, an observer south of the particle injection region measures an energy spectra with a positive slope at lower energies for a prolonged amount of time. 
The observer located in the solar equatorial plane sees the energy spectrum evolving from a simple power law to a power law with an exponential rollover at high energies due to the combined effects of drifts and adiabatic deceleration (see also the middle panel of Fig.~\ref{fig:3obsImf700}).

An evident way of reproducing the SEP flood phenomenon is to have strong spatial diffusion \citep{mcKibben72} as diffusion tends to mitigate any gradients present in a physical system. 
If cross-field diffusion should be responsible for the reservoir phenomenon, then particle drifts impose a lower limit on the strength of the needed cross-field diffusion. This is because the cross-field diffusion needs to be strong enough to mitigate any asymmetry introduced by the particle drifts. 
The cross-field diffusion used in Section~\ref{sec:cfd} ($\alpha = 10^{-4}$) was not strong enough to wash away the drift effects on the particle distribution function. 
For a cross-field diffusion ten times stronger,  ($\alpha = 10^{-3}$), the asymmetry between the intensities in the northern and southern hemispheres is diminished. 
However, the efficiency of cross-field diffusion to mitigate the effect of particle drifts might strongly depend on the properties of the diffusion tensor as, for example, its dependence on the particle energy. 
Hence more simulations with different diffusion models are required to clarify this point. 

Finally, the reservoir phenomenon has been explained in the past as the result of particle trapping in an expanding magnetic bottle \citep{reames96}. 
In Section~\ref{sec:cir}, we  use a solar wind configuration with a CIR that naturally produces a magnetic bottle as well, although it is not expanding. 
When particle drifts were not included in the spatial part of the FTE, we observed a very slow decay in the different  energy channels, but with a slightly different decay rate.

When including the effect of drifts in the simulations, the particle intensity-time profiles were strongly modified. 
In contrast to the cases in a Parker IMF, all energy channels are now affected by the particle drifts, even the lower ones. 
The reason is that particle drifts are strongly enhanced at the forward shock wave that bounds the CIR. 
The gradient drift introduced by this forward compression wave is mostly southward in the solar equatorial plane, hence, intensifying the drift present in a nominal Parker spiral of positive polarity. 
As a result, the asymmetry of the particle distribution as a function of latitude is strongly enhanced. 

Although it is not modelled here, a magnetic bottle configuration behind a CME is also likely to be characterised by strong magnetic field gradients and, hence, strong particle drifts. 
Any model attempting to explain the reservoir phenomenon should take this into account, although significant cross-field diffusion might diminish or remove the effect of drifts. 
Full-orbit particle simulations indicate that the particle drift terms as appearing in the isotropic Parker equation can be quenched when strong turbulence is present \citep[see e.g.][and references therein]{engelbrecht19}. However, the results of these full-orbit simulations might   be altered strongly near high amplitude compression waves or shock waves, where strong magnetic field gradients amplify the particle drifts. In addition, the level of turbulence near shock waves depends on the shock geometry, since, for example, the excitation of Alfv\'en waves due to particle streaming is proportional to the cosine of the angle between the upstream magnetic field and the shock normal direction \cite[e.g.][ and references therein]{tylka05,tylka06}. Quasi-perpendicular shocks, such as the forward shock in our simulated CIR, might thus be characterised by reduced levels of turbulence such that the particle drifts are not quenched. 

Finally, it is worth noting that the proportionality of drifts with energy will increase their effects on high-energy particles \citep[e.g.][]{marsh13} and that the solar wind often contains compression and rarefaction regions, as well as current sheets which modify the strength and the direction of drifts. 
Hence, more studies are required that include the effect of more realistic non-nominal solar wind configurations in order to better understand the effects that drifts might have on the observed  intensity-time profiles of SEP events. In this respect, Parker Solar Probe and the upcoming Solar Orbiter observations may help disentangle all these effects. 

\begin{acknowledgements}
N.\ Wijsen is supported by a PhD Fellowship of the Research Foundation Flanders (FWO). The computational resources and services used in this work were provided by the VSC (Flemish Supercomputer Center), funded by the Research Foundation Flanders (FWO) and the Flemish Government – department EWI. The work at University of Barcelona was partly supported by the Spanish Ministry of Economy, Industry and Competitivity under the project MDM-2014-0369 of ICCUB (Unidad de Excelencia ‘María de Maeztu’). The work at University of Helsinki was carried out in the Finnish Centre of Excellence in Research of Sustainable Space (Academy of Finland grant numbers 312390 and 312351). S.\ Poedts was supported by the projects GOA/2015-014 (KU Leuven), G.0A23.16N (FWO-Vlaanderen) and C~90347 (ESA Prodex).
\end{acknowledgements}

\bibliographystyle{aa.bst}
\bibliography{sep_bib}

\end{document}